%% file: main.tex
 \documentclass[acmsmall,screen,nonacm]{acmart}
\input{packages}

\input{latin}

\AtBeginDocument{%
  \providecommand\BibTeX{{%
    \normalfont B\kern-0.5em{\scshape i\kern-0.25em b}\kern-0.8em\TeX}}}

\begin{document}

\title{Beyond the Phase Ordering Problem: Finding the Globally Optimal Code \wrt Optimization Phases}

\author{Yu Wang}
\affiliation{%
	\institution{Nanjing University}
	\country{China}
}
\email{yuwang\_cs@smail.nju.edu.cn}

\author{Hongyu Chen}
\affiliation{%
	\institution{Nanjing University}
	\country{China}
}
\email{h.y.chen@smail.nju.edu.cn}

\author{Ke Wang}
\authornote{Corresponding author.}
\affiliation{%
	\institution{Visa Research}
	\country{USA}
}
\email{keowang@hotmail.com}


\input{abstract}




\maketitle

\input{intro}

\input{formv2}
\input{methodv2}

\input{frame}
\input{evav3}
\input{related}

\input{conclusion}

\bibliographystyle{ACM-Reference-Format}
\bibliography{reference}

\appendix

\end{document}


\title[Supplementary Material]{Supplementary Material}

\maketitle 

\section{Examples}\label{sec:xxx}

Figure~\ref{fig:example2} shows an example reduced from \textit{OpenSSL}. 
According to the semantics of the program, there are three particular cases to be handled: (1) first if \myinline{writing == -1}, the program simply \myinline{return 0}; (2) if \myinline{writing == 1}, it will first execute \myinline{(*s)(NULL)}, later \myinline{*s = NULL}, and finally \myinline{return 0}; (3) for any other value of \myinline{writing}, \myinline{*s = NULL} and \myinline{return 0} will be executed. The reason that LLVM, even with exhaustive search, can not recognize the semantics of the program is due to the \myinline{if} condition at line 15 \myinline{if (writing == -1 || ret == -3)} in Figure~\ref{fig:example2:mut}. After we reduce the \myinline{if} statement from line 15-18 in Figure~\ref{fig:example2:mut} into the code construct from line 14-22 in Figure~\ref{fig:moti:ori}, LLVM can reason that only when \myinline{writing == -1}, \myinline{return 0} is the only statement that gets executed. Like any other value except \myinline{1}, \myinline{writing == 3} makes \myinline{*s = NULL} get executed. Finally, if \myinline{writing == 1}, \myinline{(*s)(NULL)} will be executed first.

\begin{figure*}[t]
    \centering
    \begin{subfigure}{0.505\linewidth}
        \begin{subfigure}{\linewidth}
            \lstset{style=mystyle, numbers = left, xleftmargin=2em,frame=single,framexleftmargin=1.5em,basicstyle=\small\ttfamily\bfseries}
\begin{lstlisting}[morekeywords={Object, String, bool},basewidth=0.52em]
int mutant(int (**s)(void*),
           int writing) {
  bool or_cond;
  int ret = writing;
  if (writing == -3) {
    if (false)
      // noop
    else
      *s = NULL;
  } else {
    if (writing == 1)
      (*s)(NULL);
  }
  int sub = ret - (-3);
  if (writing == -1) 
    or_cond = true;
  else
    or_cond = sub == 0
  if (or_cond)
    // noop
  else
    *s = NULL;
  return 0;
}
\end{lstlisting}
        \end{subfigure}

        \vspace*{-4pt}
        \quad \quad\quad\quad\quad \quad\quad\quad\quad $\boldsymbol{\downarrow}$ \textit{\texttt{compiled to}} \\
        \vspace*{-4pt}

        \begin{subfigure}{\linewidth}
            \lstset{style=mystyle, numbers = left, xleftmargin=2em,frame=single,framexleftmargin=1.5em,basicstyle=\small\ttfamily\bfseries}
\begin{lstlisting}[morekeywords={Object, String, bool},basewidth=0.52em]
int mutant(int (**s)(void*), 
           int writing) {
  switch (writing) {
    case 1:
      (*s)(NULL);
      // fall through
    default:
      *s = NULL;
      // fall through

    case -1:
      return 0;
      
  }
}
\end{lstlisting}

        \end{subfigure}
        \caption{}
        \label{fig:moti:ori}
    \end{subfigure}
    \vspace*{-5pt}
    \,
    \begin{subfigure}{0.475\linewidth}
        \begin{subfigure}{\linewidth}
            \lstset{style=mystyle, numbers = left, xleftmargin=2em,frame=single,framexleftmargin=1.5em,basicstyle=\small\ttfamily\bfseries}
\begin{lstlisting}[morekeywords={Object, String, bool},basewidth=0.52em]
int src(int (**s)(void*),
        int writing) {
    
  int ret = writing;
  if (writing == -3) {
      if (false)
        // noop
      else
        *s = NULL;
  } else {
      if (writing == 1) 
        (*s)(NULL);
  }
    
  if (writing == -1 || ret == -3)
    return 0;
  else
    *s = NULL;

  return 0;
}
\end{lstlisting}
        \end{subfigure}

        \vspace*{-4pt}
        \quad \quad\quad\quad\quad \quad\quad\quad\quad $\boldsymbol{\downarrow}$ \textit{\texttt{compiled to}} \\
        \vspace*{-4pt}

        \begin{subfigure}{\linewidth}
            \lstset{style=mystyle, numbers = left, xleftmargin=2em,frame=single,framexleftmargin=1.5em,basicstyle=\small\ttfamily\bfseries}
\begin{lstlisting}[morekeywords={Object, String, bool},basewidth=0.52em]
int src(int (**s)(void*), 
        int writing) {
  switch (writing) {
    case 1:
        (*s)(NULL);
        "goto case -3";
    default:
        if ((writing & (-3)) == -3)
            return 0;
        else
            "goto case -3";
    case -3:
        *s = NULL;
        return 0;
  }
}
\end{lstlisting}
        \end{subfigure}

        \caption{}
        \label{fig:example2:mut}
    \end{subfigure}
    \caption{}
    \label{fig:example2}
\end{figure*}

\clearpage

Figure~\ref{fig:example4} gives another example, whose code is reduced from QEMU. This example mostly relates to how the compiler deals with folding branches. In \myinline{src}, the compiler fetches the identical operator the operand \myinline{& 1} from the two branches at first. To collapse the conditional branch, the compiler fetches \myinline{rgb >> 8} and \myinline{rgb >> 16} separately, producing the conditional move \myinline{shr1_sink_v = cmp ? 8 : 16} and a shift \myinline{shr1_sink = rgb >> shr1_sink_v}. Similarly, the compiler fetches \myinline{rgb} and \myinline{shr = rgb >> 8} separately, producing the conditional move \myinline{shr1_sink_v = cmp ? rgb : shr}. Finally, these conditional moves are combined into corresponding arithmetic expression \myinline{shr1_sink | shr2_sink}. However, \tool reverses \textit{Reassociate}, reordering \myinline{(rgb >> 16) | (rgb >> 8)} to \myinline{(rgb >> 8) | (rgb >> 16)}. Then the optimization procedure is done as \myinline{src} is done: (1) sink identical \myinline{ & 1} to the \myinline{return} statement. (2) sink left-handsided operand of bitwise-or \myinline{rgb >> 8} and \myinline{rgb >> 8} of two branches to \myinline{cmp ? rgb >> 8 : rgb >> 8}, which would be folded \myinline{rgb >> 8}. (3) sink right-handsided operand of bitwise-or \myinline{rgb} and \myinline{rgb >> 16} of two branches to \myinline{cmp ? rgb : rgb >> 16}. Compared with \myinline{src}, \myinline{mutant} saves a select/conditional-move instruction.

\begin{figure*}[t]
    \centering
    \begin{subfigure}{0.505\linewidth}
        \begin{subfigure}{\linewidth}
            \lstset{style=mystyle, numbers = left, xleftmargin=2em,frame=single,framexleftmargin=1.5em,basicstyle=\footnotesize\ttfamily\bfseries}
\begin{lstlisting}[morekeywords={Object, String, bool},basewidth=0.52em]
unsigned mutant(unsigned rgb,
                unsigned bpp) {
  unsigned ret = 0;
  if (bpp - 0 == 0) {
    ret = ((rgb >> 8) | rgb) & 1;
  } else {
    ret = ((rgb >> 8) | (rgb >> 16)) & 1;
  }
  return ret;
}
\end{lstlisting}
        \end{subfigure}

        \vspace*{-4pt}
        \quad\quad\quad\quad \quad\quad\quad\quad \quad\quad\quad\quad $\boldsymbol{\downarrow}$ \textit{\texttt{compiled to}} \\
        \vspace*{-4pt}

        \begin{subfigure}{\linewidth}
            \lstset{style=mystyle, numbers = left, xleftmargin=2em,frame=single,framexleftmargin=1.5em,basicstyle=\footnotesize\ttfamily\bfseries}
\begin{lstlisting}[morekeywords={Object, String, bool},basewidth=0.52em]
unsigned mutant(unsigned rgb,
                unsigned bpp) {
  bool cmp = bpp == 0;
  unsigned shr1 = rgb >> 16;
  unsigned shr1_sink = cmp?rgb:shr1;
  unsigned shr2 = rgb >> 8;
    
  return (shr2 | shr1_sink) & 1;
}
\end{lstlisting}

        \end{subfigure}
        \caption{}
        \label{fig:moti:ori2}
    \end{subfigure}
    \vspace*{-5pt}
    \,
    \begin{subfigure}{0.475\linewidth}
        \begin{subfigure}{\linewidth}
            \lstset{style=mystyle, numbers = left, xleftmargin=2em,frame=single,framexleftmargin=1.5em,basicstyle=\footnotesize\ttfamily\bfseries}
\begin{lstlisting}[morekeywords={Object, String, bool},basewidth=0.52em]
unsigned src(unsigned rgb,
             unsigned bpp) {
  unsigned ret = 0;
  if (bpp == 0) {
    ret = ((rgb >> 8) | rgb) & 1;
  } else {
    ret = ((rgb >> 16) | (rgb >> 8)) & 1;
  }
  return ret;
}
\end{lstlisting}
        \end{subfigure}

        \vspace*{-4pt}
        \quad\quad\quad\quad \quad\quad\quad\quad \quad\quad\quad\quad $\boldsymbol{\downarrow}$ \textit{\texttt{compiled to}} \\
        \vspace*{-4pt}

        \begin{subfigure}{\linewidth}
            \lstset{style=mystyle, numbers = left, xleftmargin=2em,frame=single,framexleftmargin=1.5em,basicstyle=\footnotesize\ttfamily\bfseries}
\begin{lstlisting}[morekeywords={Object, String, bool},basewidth=0.52em]
unsigned src(unsigned rgb,
             unsigned bpp) {
  bool cmp = bpp == 0;
  unsigned shr = rgb >> 8;
  unsigned shr2_sink = cmp?rgb:shr;
  unsigned shr1_sink_v = cmp ? 8 : 16;
  unsigned shr1_sink = rgb >> shr1_sink_v;
  return (shr1_sink | shr2_sink) & 1;
}
\end{lstlisting}
        \end{subfigure}

        \caption{}
        \label{fig:example4:mut}
    \end{subfigure}
    \caption{}
    \label{fig:example4}
\end{figure*}

\newpage

\section{A List of Program Used in the Evaluation of IBO}

\begin{itemize}
    \item Linux
    \item Redis
    \item Zstandard
    \item NumPy
    \item Protobuf 
    \item FFmpeg 
    \item OpenSSL
    \item z3
    \item PostgreSQL
    \item Ruby 
    \item PHP 
    \item QEMU     
\end{itemize}

%% file: packages.tex
\usepackage{xcolor}
\usepackage{enumitem}
\usepackage{color}
\usepackage{adjustbox}
\usepackage{array}
\usepackage{tabularx} 
\usepackage{xspace}
\usepackage{pgf-pie}
\usepackage{bm}
\usepackage{stackrel}

\usepackage{booktabs}
\usepackage{subcaption}
\usepackage{listings} 
\usepackage{lstautogobble}
\usepackage{tikz}
\usepackage{caption}
\usepackage{pict2e}
\usepackage{algorithm}
\usepackage{multirow}
\usepackage{mathtools}	
\usepackage{mdframed}
\usepackage{syntax}
\usepackage[skins]{tcolorbox}
\usepackage{pgfplots}
\pgfplotsset{
	compat=1.4,
	small,
	legend style={
		at={(0.99,0.99)},
		anchor=north east,
		font=\bfseries,
	},
	label style={font=\sffamily\small\bfseries}
}
\usepackage{diagbox}
\usepackage[normalem]{ulem}
\let\UnderScore_

\newcommand\redout{\bgroup\markoverwith
	{\textcolor{red}{\rule[.45ex]{1.5pt}{1.pt}}}\ULon}

\global\mdfdefinestyle{default}{%
	usetwoside=false,%
	linecolor=black,linewidth=1pt,%
	innerleftmargin=5pt,innerrightmargin=5,%
	everyline=true
}

\newcommand{\tabincell}[2]{\begin{tabular}{@{}#1@{}}#2\end{tabular}}

\theoremstyle{definition}
\newtheorem{definition}{Definition}[section]
\newtheorem{theorem}{Theorem}[section]

\makeatletter
\newenvironment{btHighlight}[1][]
{\begingroup\tikzset{bt@Highlight@par/.style={#1}}\begin{lrbox}{\@tempboxa}}
	{\end{lrbox}\bt@HL@box[bt@Highlight@par]{\@tempboxa}\endgroup}

\newcommand\btHL[1][]{%
	\begin{btHighlight}[#1]\bgroup\aftergroup\bt@HL@endenv%
	}
	\def\bt@HL@endenv{%
	\end{btHighlight}%
	\egroup
}
\newcommand{\bt@HL@box}[2][]{%
	\tikz[#1]{%
		\pgfpathrectangle{\pgfpoint{1pt}{0pt}}{\pgfpoint{\wd #2}{\ht #2}}%
		\pgfusepath{use as bounding box}%
		\node[anchor=base west, fill=orange!25,outer sep=.5pt,inner xsep=0.5pt, inner ysep=0.15pt, rounded corners=1pt, minimum height=\ht\strutbox-.1pt,#1]{\raisebox{.01pt}{\strut}\strut\usebox{#2}};
	}%
}
\newcounter{procedurealgorithm}

\definecolor{add-line}{rgb}{0.84, 0.89, 0.97}
\definecolor{add-word}{rgb}{0.5, 0.7, 0.98}
\definecolor{ForestGreen}{RGB}{63,147,88}
\definecolor{remove-line}{rgb}{1.0, 0.93, 0.73}
\definecolor{remove-word}{rgb}{1.0, 0.8, 0.2}
\definecolor{Gray}{gray}{0.9}
\definecolor{orange}{rgb}{0.84, 0.89, 0.97}

\lstdefinestyle{mystyle}{  
	frame=single, 
	framexleftmargin=0pt,
	commentstyle=\color{ForestGreen},
	keywordstyle=\color{blue}\bfseries,
	numberstyle=\normalsize\color{gray},
	stringstyle=\color{purple},
	basicstyle=\normalsize\ttfamily\bfseries,
	breakatwhitespace=false,         
	breaklines=false,                 
	captionpos=b,                    
	keepspaces=true,     
	numbers=none,                    
	numbersep=4pt,               
	showspaces=false,                
	showstringspaces=false,
	showtabs=false,                  
	tabsize=2,  
	language=C++,  
	escapechar=\@,  
	moredelim=**[is][{\btHL[fill=red!40]}]{@}{@},
	moredelim=**[is][{\btHL[fill=remove-line]}]{|*}{*|},
	moredelim=**[is][{\btHL[fill=remove-word]}]{|^}{^|},
	moredelim=**[is][{\btHL[fill=add-line]}]{||^}{^||},	
	moredelim=**[is][{\btHL[fill=add-word]}]{||*}{*||}
} 
\let\origthelstnumber\thelstnumber
\makeatletter
\newcommand*\Suppressnumber{%
  \lst@AddToHook{OnNewLine}{%
    \let\thelstnumber\relax%
     \advance\c@lstnumber-\@ne\relax%
    }%
}

\lstdefinestyle{inlinestyle}{
	language=C++,
    basicstyle=\small\ttfamily\bfseries,
}

\newcommand*\Reactivatenumber[1]{%
  \setcounter{lstnumber}{\numexpr#1-1\relax}
  \lst@AddToHook{OnNewLine}{%
   \let\thelstnumber\origthelstnumber%
   \refstepcounter{lstnumber}%
  }%
}
\makeatother

\usepackage{algpseudocode}
\renewcommand*\Call[2]{\textproc{#1}(#2)}
\algnewcommand{\LineComment}[1]{\Statex #1}
\algrenewcommand\algorithmicindent{1.0em}
\algdef{SE}[DOWHILE]{Do}{doWhile}{\algorithmicdo}[1]{\algorithmicwhile\ #1}%

\usetikzlibrary{shapes,positioning,arrows.meta, calc, decorations.pathreplacing, matrix}
\usepackage{colortbl}
\definecolor{Gray}{gray}{0.85}
\tikzset{draw half paths/.style 2 args={%
  decoration={show path construction,
    lineto code={
      \draw [#1] (\tikzinputsegmentfirst) -- 
         ($(\tikzinputsegmentfirst)!0.5!(\tikzinputsegmentlast)$);
      \draw [#2] ($(\tikzinputsegmentfirst)!0.5!(\tikzinputsegmentlast)$)
        -- (\tikzinputsegmentlast);
    }
  }, decorate
}}
\usepackage{tikz-qtree}

\newcommand{\textmtte}[1]{{\fontsize{9}{11}\fontfamily{txtt}\selectfont #1}}

\renewcommand{\texttt}[1]{\textmtte{#1}}
\newcommand{\myinline}{\lstinline[style=inlineStyle]}
\DeclareTextFontCommand{\mytexttt}{\small\ttfamily\bfseries}

\newcommand{\tool}{IBO\xspace}
\newcommand{\cmp}{\textsf{Cop}\xspace}

\newcommand{\FIBO}{IBO\xspace}

\newcommand{\poc}{\ensuremath{P_c}\xspace}
\newcommand{\pd}{\ensuremath{P_o'}\xspace}
\newcommand{\pdc}{\ensuremath{P_c'}\xspace}

\newtcolorbox{greytcolorbox}{before=\par\smallskip\centering,after=\par, before skip balanced=5pt,after skip balanced=6pt,hbox}

\newtcolorbox{whitetcolorbox}{opacityback=0,enhanced jigsaw, before=\par\smallskip\centering,after=\par, before skip balanced=5pt,after skip balanced=6pt,hbox,boxrule=1pt}

\newcommand{\phiequiv}{\mspace{-2mu}\mathrel{\ensuremath{\vcenter{\hbox{\ensuremath{\scriptscriptstyle \Phi}}} \mspace{-5mu}\equiv}}}

\newcommand{\phiLequiv}{\mspace{-3mu}\mathrel{\ensuremath{\vcenter{\hbox{\ensuremath{\scriptscriptstyle \Phi_{\mkern-2mu\scalebox{.6}{$\scriptscriptstyle L$}}}}} \mspace{-6.2mu}\equiv}}}

\newcommand{\sep}{\hbox{\textsf{SEP}}\xspace}

%% file: latin.tex
\newcommand{\eg}{\hbox{\emph{e.g.}}\xspace}
\newcommand{\ie}{\hbox{\emph{i.e.}}\xspace}

\newcommand{\wrt}{\hbox{\emph{w.r.t.}}\xspace}

\newcommand{\vice}{\hbox{vice versa}\xspace}

%% file: abstract.tex
\begin{abstract}

In this paper, we propose a new concept called \textit{semantically equivalence} \wrt \textit{optimization phases} \textit{(\sep)}, which defines the set of programs a compiler considers semantically equivalent to the input using a set of optimization phases. We show both theoretically and empirically that solving the phase ordering problem does not necessarily result in the most efficient code among all programs that a compiler deems semantically equivalent to the input, hereinafter referred to as the global optimal code \wrt optimization phases. 

To find the global optimal code \wrt optimization phases, we present a conceptual framework, leveraging the reverse of existing optimization phases. In theory, we prove that the framework is capable of finding the global optimal code for any program. We realize this framework into a technique, called \textit{iterative bi-directional optimization (\tool)}, which performs both the normal and reverse optimizations to increase and decrease the efficiency of the generated code, respectively.

We evaluate \tool on C/C++ files randomly extracted from highly mature and influential programs (\eg, Linux kernel, OpenSSL, Z3). Results show that \tool frequently generates more efficient code --- measured by either code size or runtime performance --- than exhaustive search, which is the solution to the phase ordering problem. We also find by simply incorporating \tool's reverse optimization phases, the effectiveness of the optimization of state-of-the-art compilers (\eg, GCC/LLVM) can be significantly improved.

\end{abstract}

%% file: intro.tex

\section{Introduction}
\label{sec:intro}

The phase ordering problem has been a long-standing issue in compiler optimizations. The order in which optimization phases are applied to a program can result in different code, with potentially significant efficiency variations among them~\cite{po1,po2,po3}. Therefore, finding the optimal sequence of optimization phases is crucial, especially in performance-critical domains. The past four decades has witnessed a tremendous amount of effort devoted to solving the phase ordering problem~\cite{phaseordering10,phaseordering11,phaseordering9,kulkarni2009practical,AutoPhase,ashouri2017micomp}, with some achieving remarkable success.

One aspect of compiler optimizations that is rarely discussed in the compiler literature is the notion of semantic equivalence from compilers' perspective. This is an important issue, because, for one, the notion defines the set of programs that a compiler should consider when optimizing an input program (\ie, all programs the compiler deems semantically equivalent to the input). In addition, this notion clearly differs from the standard definition of semantic equivalence based on the formal semantics of programming languages. Because compilers cannot reason about the semantics of a program in a general sense that program verification tools (\eg, theorem provers, SMT solvers) can. For example, no existing compilers optimize bubble sort into quick sort, which implies that they do not recognize the semantic equivalence between the two sorting routines.
Given the notion of semantic equivalence from compiler's standpoint, one may naturally expect them to aspire to generate the most efficient code among all programs they deem semantically equivalent to the input. Although lacking a formal argument, solving the phase ordering problem is understood in some way or another for achieving this aspiration, especially given the decades of effort and dedication invested. As the first main objective of this paper, we thoroughly evaluate whether applying the optimal ordering of a compiler's optimization phases produces the most efficient code among all programs it recognizes as semantically equivalent to the input.

\begin{figure*}[t]
    \centering
    \captionsetup{skip=5pt}
    \begin{subfigure}{0.5\linewidth}
        \centering
        \captionsetup{skip=0pt}
        \lstset{style=mystyle, basicstyle=\footnotesize\ttfamily\bfseries}
\begin{lstlisting}[morekeywords={Object, String, bool},basewidth=0.52em]
unsigned char original(unsigned val) {
  return ((val/10) << 4) + val % 10;
}

\end{lstlisting}
        \caption{$P_a$}
        \label{fig:ori1}
        \vspace{3pt}
    \end{subfigure}

    \begin{subfigure}{0.49\linewidth}
        \centering
        \captionsetup{skip=0pt}
        \lstset{style=mystyle, basicstyle=\footnotesize\ttfamily\bfseries}
\begin{lstlisting}[morekeywords={Object, String, bool},basewidth=0.52em]
unsigned char compiledA(unsigned val) {
  return ((val/10) << 4) | val % 10;
}
\end{lstlisting}
        \caption{$P_b$}
        \label{fig:comA1}
    \end{subfigure}
    \,\,
    \begin{subfigure}{0.49\linewidth}
        \centering
        \captionsetup{skip=0pt}
        \lstset{style=mystyle, basicstyle=\footnotesize\ttfamily\bfseries}
\begin{lstlisting}[morekeywords={Object, String, bool},basewidth=0.52em]
unsigned char compiledB(unsigned val){
  return ((val/10) * 6) + val;
}
\end{lstlisting}
        \caption{$P_c$}
        \label{fig:comB1}
    \end{subfigure}

    \caption{
    (\subref{fig:ori1}) shows an input program $P_a$.~(\subref{fig:comA1}) shows the code $P_b$ that LLVM compiles $P_a$ into using the optimal ordering of optimization passes. Since the result of \lstinline[language=C++,basicstyle=\footnotesize\ttfamily\bfseries]{val \% 10}    
    occupies four bits, and the result of \lstinline[language=C++,basicstyle=\footnotesize\ttfamily\bfseries]{(val / 10) << 4} vacates the four right most bits, the bitwise or operator in $P_b$ achieves the same effect as the addition operator in $P_a$ but runs faster.~(\subref{fig:comB1}) shows another piece of compiled code $P_c$ related to $P_a$ and $P_b$. For illustration purpose, all compiled code are displayed at the source level.
    }
    \vspace{-4pt}
    \label{fig:moti1}
\end{figure*}

First and foremost, we formalize the notion of ``semantic equivalence from compilers' perspective'', in particular, we propose a novel concept of \textit{\underline{s}emantically \underline{e}quivalence w.r.t. optimization \underline{p}hases} \textit{(\sep)}. \sep defines for a compiler with a set of optimization phases/passes $\Phi$, whether two programs are considered semantically equivalent. Here we only provide the intuition behind \sep, and defer its formal definition to Section~\ref{sec:form}. Given two programs $P_1$ and $P_2$, if the compiler can optimize $P_1$ into $P_2$ or \vice using $\Phi$, it can determine that $P_1$ and $P_2$ are semantically equivalent, denoted by $P_1 \phiequiv P_2$ (or $P_2 \phiequiv P_1$). In light of \sep, we find, through theoretical and empirical means, applying the optimal ordering of optimization phases $\Phi$ to a program $P$ may not result in the most efficient program that shares the same semantics as $P$ \wrt $\Phi$. In other words, solving the phase ordering problem, in fact, does not achieve compilers' aspiration for generating the most efficient code. To prove this point, we provide a concrete example. Figure~\ref{fig:ori1} depicts a function extracted from Linux kernel\footnote{\url{https://github.com/torvalds/linux/blob/master/lib/bcd.c\#L11-L14}}, which translates binary to BCD (binary-coded decimal) code. We denote this function by $P_a$. Figure~\ref{fig:comA1} shows the code (denoted by $P_b$) that LLVM compiles $P_a$ into. All compiled code are displayed at the source level for illustration purpose. It is worth mentioning that LLVM engineers have verified that $P_b$ is the result of compiling $P_a$ with the optimal ordering of all optimization passes in LLVM. In particular, they take a brute-force approach which exhaustively enumerates all possible orderings of LLVM's optimization passes, and find $P_b$ is the most efficient (according to either code size or expected runtime) that LLVM can produce. In the meanwhile, Figure~\ref{fig:comB1}, denoted by $P_c$, shows another piece of code for which we make two critical observations:

\begin{itemize}[leftmargin=*]
    \item \textit{\textbf{\boldsymbol{$P_c$} exhibits higher efficiency than \boldsymbol{$P_b$}.}} In terms of code size, two state-of-the-art compilers --- LLVM and GCC --- produce four fewer instructions on X86 and two fewer instructions on RISC-V for $P_c$. Additionally, $P_c$ is estimated to run nearly four times faster on X86 and 10\% faster on RISC-V according to the LLVM machine code analyzer (MCA), which is generally regarded as an accurate and reliable performance analyzer for straight-line machine code.

    \item \boldsymbol{$P_c \phiLequiv{} P_b$}
    \textit{\textbf{(}}\boldsymbol{{$\Phi_{\hspace{-.6pt}\scalebox{.7}{$\scriptscriptstyle L$}}$}} \textit{\textbf{in} }\textbf{``}\,\boldsymbol{$\phiLequiv{}$}\!\textbf{''}
     \textit{\textbf{represents the set of optimization passes in LLVM).}} The reason is, as illustrated in Figure~\ref{fig:moti}, there exists another program $P_d$ in Figure~\ref{fig:reverse}, which can be optimized into $P_b$ and $P_c$ respectively. According to the definition of 
    \textsf{SEP}, we have $P_d \phiLequiv P_b$ and $P_d \phiLequiv P_c$. Due to the transitivity of \textsf{SEP}, we have $P_c \phiLequiv P_b$.     

\end{itemize}

Because $P_b$ is the result of applying the optimal ordering of 
{$\Phi_{\hspace{-.6pt}\scalebox{.7}{$\scriptscriptstyle L$}}$} to $P_a$, and $P_c$ is both more efficient than $P_b$ and semantically equivalent to $P_a$ \wrt \,\!\!\! {$\Phi_{\hspace{-.6pt}\scalebox{.7}{$\scriptscriptstyle L$}}$}, this example demonstrates that solving the ordering problem for a given set of optimization passes $\Phi$ may not result in the most efficient code that is semantically equivalent to the input \wrt $\Phi$. It is important to emphasize that this issue should not be viewed as the deficiency in the solution to the phase ordering problem, but rather as a consequence of the inadequate definition of the phase ordering problem from the outset. Specifically, phase ordering problem submits to the constraint imposed by optimization passes, that is, all of them are designed to monotonically increase the efficiency of input programs.\footnote{
Some optimizations do not strictly increase the efficiency of input programs, such as reassociation (\eg, \lstinline[language=C++,basicstyle=\footnotesize\ttfamily\bfseries]{a+b} $\rightarrow$ 
\lstinline[language=C++,basicstyle=\footnotesize\ttfamily\bfseries]{b+a}). However, they are only performed as an intermediate step to facilitate subsequent core optimizations that do improve efficiency. Thus, from the core optimizations standpoint, output programs are always more efficient than input programs.} As a result, less efficient programs yet denoting the same semantics as the input are ignored. When such less efficient programs can be optimized into a more efficient program than the original input can ever be optimized into, applying the optimal phase ordering to the input will end up with suboptimal code. 
This is precisely the issue that the example above exposes. As illustrated in Figure~\ref{fig:moti}, $P_c$ can only be obtained by optimizing $P_d$, a program that is less efficient than $P_a$ but semantically equivalent to $P_a$ \wrt LLVM's optimization passes {$\Phi_{\hspace{-.6pt}\scalebox{.7}{$\scriptscriptstyle L$}}$}. Therefore, $P_c$ is beyond what LLVM can generate for $P_a$ even if it applies the optimal ordering of {$\Phi_{\hspace{-.6pt}\scalebox{.7}{$\scriptscriptstyle L$}}$}.


\begin{figure*}[t]
    \captionsetup{skip=5pt}
    \centering
    \begin{subfigure}{0.5\linewidth}
        \centering
        \captionsetup{skip=0pt}
        \lstset{style=mystyle, basicstyle=\footnotesize\ttfamily\bfseries}
\begin{lstlisting}[morekeywords={Object, String, bool},basewidth=0.52em]
unsigned char reversed(unsigned val) {
  return ((val/10) * 16) + val - (val/10) * 10;
}
\end{lstlisting}
        \captionsetup{labelformat=empty}
        \caption{$P_d$ }
        \label{fig:reverse}
    \end{subfigure}

    \begin{center}
        \vspace*{-4pt}
        $\boldsymbol{\swarrow}$  \quad\quad\quad\quad \quad\quad\quad\quad \quad\quad\quad\quad $\boldsymbol{\searrow}$  \\
    \end{center}

    \begin{subfigure}{0.49\linewidth}
        \centering
        \captionsetup{skip=0pt}
        \lstset{style=mystyle, basicstyle=\footnotesize\ttfamily\bfseries}
\begin{lstlisting}[morekeywords={Object, String, bool},basewidth=0.52em]
unsigned char reversed_opted(unsigned val) {
  return ((val/10) << 4) + val - (val/10) * 10;
}
\end{lstlisting}
        \captionsetup{labelformat=empty}
        \caption{$P_e$}
        \label{fig:opt1a}
    \end{subfigure}
    {\phantom{$\;\boldsymbol{\rightarrow}\;\,$}}%
    \begin{subfigure}{0.42\linewidth}
        \centering
        \captionsetup{skip=0pt}
        \lstset{style=mystyle, basicstyle=\footnotesize\ttfamily\bfseries}
\begin{lstlisting}[morekeywords={Object, String, bool},basewidth=0.52em]
unsigned char compiledB(unsigned val){
  return ((val/10) * 6) + val;
}
\end{lstlisting}
        \setcounter{subfigure}{3}%
        \captionsetup{labelformat=empty}
        \caption{$P_c$}
        \label{fig:opt2}
    \end{subfigure}

    \begin{center}
        \vspace*{-4pt}
        $\boldsymbol{\downarrow}$  \quad\quad\quad\quad \quad\quad\quad \;\!\! \quad\quad\quad\quad\quad \quad\quad\quad\quad\quad\quad\quad\quad \\
    \end{center}

    \begin{subfigure}[m]{0.49\linewidth}
        \centering
        \captionsetup{skip=0pt}
        \lstset{style=mystyle, basicstyle=\footnotesize\ttfamily\bfseries}
\begin{lstlisting}[morekeywords={Object, String, bool},basewidth=0.52em]
unsigned char original(unsigned val) {
  return ((val/10) << 4) + val % 10;
}

\end{lstlisting}
        \setcounter{subfigure}{2}%
        \captionsetup{labelformat=empty}
        \caption{$P_a$ }
        \label{fig:opt1b}
    \end{subfigure}
    {\raisebox{7pt}{$\;\boldsymbol{\rightarrow}\;\,$}}%
    \begin{subfigure}[m]{0.42\linewidth}
        \centering
        \captionsetup{skip=0pt}
         \lstset{style=mystyle, basicstyle=\footnotesize\ttfamily\bfseries}
\begin{lstlisting}[morekeywords={Object, String, bool},basewidth=0.52em]
unsigned char compiledA(unsigned val) {
  return ((val/10) << 4) | val % 10;
}
\end{lstlisting}
        \setcounter{subfigure}{4}%
        \captionsetup{labelformat=empty}
        \caption{$P_b$}
        \label{fig:opt1c}
    \end{subfigure}
    \caption{
    Optimizing $P_d$ along two different paths. In the first sequence of optimizations ($P_d\rightarrow P_e\rightarrow P_a\rightarrow P_b$), 
    \lstinline[language=C++,basicstyle=\footnotesize\ttfamily\bfseries]{(val / 10) * 16} in $P_d$ is optimized into \lstinline[language=C++,basicstyle=\footnotesize\ttfamily\bfseries]{(val / 10) << 4}
    in $P_e$. Then
    \lstinline[language=C++,basicstyle=\footnotesize\ttfamily\bfseries]{(val / 10) * 10} in $P_e$ is optimized into \lstinline[language=C++,basicstyle=\footnotesize\ttfamily\bfseries]{val \% 10}
    in $P_a$. Next, the addition operator ``+'' in $P_a$ is replaced with the bitwise operator ``|'' in $P_b$. In another sequence of optimizations ($P_d\rightarrow P_c$), 
    \lstinline[language=C++,basicstyle=\footnotesize\ttfamily\bfseries]{(val / 10) * 16}
    and \lstinline[language=C++,basicstyle=\footnotesize\ttfamily\bfseries]{-(val / 10) * 10} in $P_d$ is combined into \lstinline[language=C++,basicstyle=\footnotesize\ttfamily\bfseries]{(val / 10) * 6} in $P_c$.
    }
    \vspace{-8pt}
    \label{fig:moti}
\end{figure*}

This shortcoming of the phase ordering problem reaffirms the necessity of generating the most efficient code among all programs that are considered semantically equivalent to the input by compilers. Hereinafter, we refer to this problem as \textit{finding the globally optimal code} \wrt \textit{optimization phases}, which we believe should be the ultimate goal of compilers for generating the most efficient code. Moreover, this new problem presents significant advantages from both theoretic and empirical perspectives compared to the phase ordering problem, which makes it the right problem to solve.

\begin{itemize}[leftmargin=*]
\item \textit{\textbf{Generality.}} Theoretically, the problem of generating the global optimum is a generalization of the phase ordering problem. In essence, it can be broken down into two sub-problems: finding a program $O$ that is semantically equivalent to the input $P$ \wrt a given set of optimization phases $\Phi$; and applying the optimal ordering of $\Phi$ to $O$ to generate the globally optimal code $Q$ (Theorem~\ref{theo:relation} proves why the problem can be broken down in this manner). This means, when the program $O$ is the input $P$ itself, the problem of finding the global optimum $Q$ is reduced into the phase ordering problem. 

\item \textit{\textbf{Insensitivity.}} Practically, solving the problem of finding the global optimum removes compilers' sensitivity to the form in which  code is written. Figure~\ref{fig:moti1} clearly demonstrates that code can be written in a way that makes the global optimum out of compilers' reach. If the problem is solved, the global optimum can always be obtained regardless how the original code is expressed. This can improve software performance without breaking the existing coding standards for ensuring readability and maintainability.
\end{itemize}

Similar to the phase ordering problem, the globally optimal code varies with the given set of optimization phases. This means that the problem can never be solved by merely augmenting existing optimization passes. The fundamental issue, as explained earlier, is less efficient programs are never explored during the optimization process. Despite a larger set of optimization passes, this core issue remains. As a results, compilers will continue to miss the globally optimal code.

Motivated by the shortcoming of the phase ordering problem, this paper proposes a conceptual framework for finding the globally optimal code \wrt optimization phases. Theoretically, for any program $P$, the framework is capable of finding the most efficient code $Q$ that is semantically equivalent to $P$ \wrt a given set of optimization phases $\Phi$. The key idea is to incorporate into compilers the reverse of each $\phi \in \Phi$, allowing compilers to perform bi-directional optimizations to both increase and decrease the efficiency of the generated code. The reason this framework is capable of finding the global optimum $Q$ for any program $P$ is two-fold. First, incorporating the reverse of optimization passes does not change the set of semantically equivalent programs $\mathcal{P}$ \wrt $\Phi$, since reverse optimizations cannot reach new programs outside of $\mathcal{P}$. Second, reverse optimizations, along with the original ones, enable the framework to explore any program within $\mathcal{P}$. On the other hand, not all reverse optimizations are useful. For example, the reverse of deadcode elimination, which inserts deadcode to the input program, does not serve the purpose. Therefore, the key to finding the globally optimal code $Q$ is to identify the minimum set of optimization passes to reverse, with which compilers can find $Q$ as they would have with $\Phi$. When the minimum set of reverse optimizations is identified, in the future we can focus on formulating a policy using such reverse optimizations to discover $Q$.

At the technical level, we propose \textit{iterative bi-directional optimizations (IBO)}, a method for realizing the conceptual framework introduced above. \tool performs iterations of bi-directional optimizations which consist of (1) applying the reverse of optimization passes to generate less efficient programs, and (2) optimizing each generated program using the optimal phase ordering. Clearly, \tool does not solve the problem of generating the globally optimal code. By performing a finite number of iterations of bi-directional optimizations, \tool cannot explore the entire set of programs that are semantically equivalent to the input \wrt optimization phase, and thus may only arrive at a local optimum. Nevertheless, it already generates more efficient code than the exhaustive search, which can be regarded as the solution to the phase ordering problem. Specifically, we use 15 well-established, highly influential programs (\eg, Linux Kernel, OpenSSL) for evaluation. From these 15 programs, we randomly extract 50 C/C++ files for which the exhaustive search is feasible. Results show that \tool generates more efficient code --- according to either code size or runtime performance --- for 20 files 
with only
two iterations of bi-directional optimizations. This shows the progress we can make toward generating the globally optimal code \wrt optimization phases. 

Given the absence of an efficient algorithm for solving the phase ordering problem, state-of-the-art compilers (\eg, GCC/LLVM) employ their own optimization pipelines to generate the most efficient code possible. Under this circumstance, we evaluate if our conceptual framework can also be utilized to improve the effectiveness of GCC/LLVM's optimization. Adapting to this evaluation, \tool first applies reverse 
optimizations to generate less efficient programs as usual. Then, for each generated program, \tool directly optimizes it using GCC/LLVM. Finally, we compare the code generated by \tool and GCC/LLVM. To ensure the optimal performance of GCC/LLVM, we choose their most aggressive compilation mode -O3. For a fair comparison, \tool also adopts -O3 of the respective compiler it is compared against. In this experiment, we randomly extract 200 C/C++ files from the same programs used in the previous evaluation. We find that, with only three iterations of bi-directional optimizations, \tool produces more efficient code (according to code size or runtime performance) than either GCC or LLVM for more than half of the files. These results clearly demonstrate the utility of our conceptual framework, especially its reverse optimizations, 
as they are the only distinction between \tool and GCC/LLVM's optimizations.

This paper makes the following contributions: 

\begin{itemize}
    \item We formalize the notion of semantic equivalence \wrt optimization phases, based on which we define a new problem of finding the globally optimal code \wrt optimization phases. 

    \item We prove that applying the optimal ordering of a specific set of optimization phases does not always result in the globally optimal code \wrt these optimization phases.

    \item We propose a conceptual framework capable of finding the global optimum \wrt optimization phases. At the core of this conceptual framework is the reverse of optimization passes. Furthermore, we show the problem of identifying the minimum set of optimization passes to reverse is key to finding the global optimum.

    \item We present iterative bi-directional optimizations (\tool), a method for realizing the conceptual framework, along with its evaluation. Results show that \tool frequently generates more efficient code than exhaustive search, which is deemed the solution to the phase ordering problem, and outperforms state-of-the-art industry compilers such as GCC and LLVM.
\end{itemize}

%% file: formv2.tex
\section{Formalization}
\label{sec:form}

We first provide the background of compiler optimizations and introduce the necessary terminology along the way. Then, we formalize the concept of semantic equivalence \wrt optimization phases. 

\subsection{Preliminary}
\label{subsec:pre}

For illustration purpose, we use a simple compiler, \cmp, to explain the optimization procedure. Like any typical compiler, \cmp takes a program written in a language $\mathscr{L}$
and compiles it into machine code. We only formalize the optimization component in \cmp, which is the focus of this work.

\cmp relies on a set of optimization phases (or optimization passes) $\Phi$ to optimize an input program. Each optimization pass $\phi \in \Phi$ consists of a sequence of semantically-preserving program transformations $T_\phi$ (\eg, removing deadcode, replacing variables/expressions with constants, moving loop invariants out of the loop). Here, we use the subscript to denote the optimization pass $\phi$ that $T_\phi$ belongs to. By combining the transformations from all optimization passes in \cmp, we obtain the entire set of transformations $\mathcal{T}$ that \cmp uses for optimization
\[ \mathcal{T} = \bigcup\limits_{\forall \phi \in \Phi} \mathit{set}(T_{\phi}) \]
where $\mathit{set}(\cdot)$ is a function that converts sequences to sets (by ignoring the order of elements and removing duplicates in an sequence). Given the set of all transformations $\mathcal{T}$, it follows that $T_{\phi} \in \mathcal{T}^{*}$ where $*$ denotes the Kleene star.

Each transformation $t \in \mathcal{T}$ is a function $t: \mathscr{L} \rightarrow \mathscr{L}$ that receives a program $P$ as input, and generates another program $Q$ as output, assuming \cmp's optimization operates at the source code level (\ie, in language $\mathscr{L}$). When \cmp applies an optimization pass $\phi$ to an input program, essentially, it applies $T_\phi$ to the program. This means that each individual transformation $t_i$ in $T_\phi$ is successively applied to the result of the previous transformation. For example, given a program $P$ and $T_\phi=[t_1,t_2,\dots,t_n]$, the application of $T_\phi$ to $P$ is defined as $T_\phi(P) = t_n(\dots t_2(t_{1}(P)))$. Similarly, applying multiple optimization passes $\varphi=[\phi_1,\phi_2,\dots,\phi_n]$ to $P$ can be formalized into $\varphi(P) = T_{\phi_{n}}\dots(T_{\phi_{2}}(T_{\phi_{1}}(P)))$.

We use a function $\delta:\mathscr{L} \rightarrow \mathbb{R}_{>0}$ ($\mathbb{R}_{>0}$ denotes the set of positive real numbers) to measure the efficiency of a program. Specifically, $\delta(P_1) > \delta(P_2)$ means $P_1$ is more efficient than $P_2$. In principle, function $\delta(\cdot)$ could depend upon code size, expected runtime, number of disk accesses, power consumption, or any other measure of resource usage. In this paper, we adopt two of the most widely applied metrics --- code size and runtime performance --- as implementations of $\delta(\cdot)$. 

Finally, we formalize the phase ordering problem in \cmp. Given a program $P$ to optimize, the goal is to find an ordering of optimization passes $\varphi \in \Phi^*$, such that applying $\varphi$ to $P$ results in the most efficient code. Formally, we seek an ordering $\varphi$ where no other ordering  $\varphi^\prime \in \Phi^*$ produces a more efficient program
$\nexists \varphi^{\prime}\! \in \Phi^* \, \delta(\varphi^{\prime}(P)) > \delta(\varphi(P))$.

\subsection{Semantic Equivalence \wrt Optimization Phases}

We now define \textit{semantic equivalence w.r.t optimization phases (\sep)}, which is central to the problem of finding the globally optimal code \wrt optimization phases.

\begin{definition}[Semantic Equivalence \wrt Optimization Phases]
	\label{def:equi}
    Two programs $P_1$ and $P_2$ are semantically equivalent \wrt a set of optimization passes $\Phi$, denoted by $P_1 \phiequiv P_2$ (or $P_2 \phiequiv P_1$), if $P_1$ can be transformed into $P_2$ or \vice by an optimization pass $\phi \in \Phi$.
\end{definition}

Since \sep is fundamentally an equivalence relation, it is transitive, therefore, we can compute the transitive closure of \sep on the set of semantically equivalent programs \wrt optimization phases. In this way, Definition~\ref{def:equi} can be extended to be: if $P_1$ can be transformed into $P_2$ or \vice via a sequence of optimization passes $\varphi \in \Phi^{*}$, then $P_1 \phiequiv P_2$.

Based on the definition of \sep, we can compute the set of all semantically equivalent programs to $P$ with $\Phi$, denoted by $\mathcal{P} = \{P^\prime|P^\prime \phiequiv P\}$.

%% file: methodv2.tex
\section{A New Problem Definition --- Finding the Globally Optimal Code \textit{w.r.t.} Optimization Phases}
\label{sec:problem}

In this section, we first define the problem of finding globally optimal code \wrt optimization phases. Then, we formally prove that solving the phase ordering problem is not the same as generating the globally optimal code \wrt optimization phases. Next, we discuss the relationship between the problem of finding the globally optimal code \wrt optimization phases and the phase ordering problem. Finally, we show that the problem is undecidable.


\subsection{Defining the Problem of Finding the Global Optimum \wrt Optimization Phases}
\label{subsec:problem}


\begin{definition}[The Globally Optimal Code \wrt Optimization Phases]
	\label{def:go}
    A program $Q$ is the global optimum of another program $P$ \wrt a set of optimization passes $\Phi$, \textit{iff} $Q$ is the most efficient among all programs that are semantically equivalent to $P$ \wrt $\Phi$ according to $\delta(\cdot)$. Formally, $\nexists P^\prime \in \mathcal{P} \, \delta(P^\prime) > \delta(Q)$ where $\mathcal{P}=\{P^\prime|P^\prime \phiequiv P\}$.
\end{definition}

The globally optimal code $Q$ for the program $P$ \wrt $\Phi$ always exists if $\delta(\cdot)$ depends on the code size only. If $\delta(\cdot)$ also takes into account the runtime performance, the global optimum only exists for programs without loops/recursions.  Therefore, to prove the existence of the global optimum when the runtime performance is a factor, we assume the bounded code size, a practical constraint that compilers also respect in the real-world. 

\begin{theorem}[Existence of The Global Optimum \wrt Optimization Phases]
\label{theo:exi}
The global optimum $Q$ for a program $P$ \wrt a set of optimization passes $\Phi$ always exists if $\delta(\cdot)$ depends on exclusively the code size. When $\delta(\cdot)$ also takes into account runtime performance, the global optimum $Q$ exists when no program can exceed a limit on the code size.
\end{theorem}
\begin{proof}
We first consider the case of $\delta(\cdot)$ depending on code size only. Because for any program $P$, no optimization can be applied for an infinite number of times, and at the same time keep reducing the code size of $P$. Therefore, there exists a program $Q$ (\ie, $Q = \varphi(P)$ where $\varphi \in \Phi^{*}$) whose size can not be reduced any further. Therefore, the global optimum of $P$ exists.

Even though there exists optimizations that can be applied to a program for an infinite number of times (\eg, loop unrolling, function inlining (for optimizing programs with recursions)), they always increase the size of the generated code. Therefore, assuming bounded code size, no optimization can be applied indefinitely and keeps improving program's runtime performance. Therefore, the global optimum for $P$ \wrt $\Phi$ exists.
\end{proof}

\subsection{Solving the Phase Ordering Problem $\ne$ Finding the Globally Optimal Code \wrt Optimization Phases}%
\label{subsec:defi}

Given the new problem definition presented in Section~\ref{subsec:problem}, we prove that solving the phase ordering problem may not result in the global optimum \wrt optimization phases.

\begin{figure}[t]
	\captionsetup{skip=5pt}
	\centering
	\begin{subfigure}{0.49\textwidth}
		\captionsetup{skip=1pt}
		\centering
		\includegraphics[trim={0pt 14pt 0 0},clip, width=0.95\linewidth]{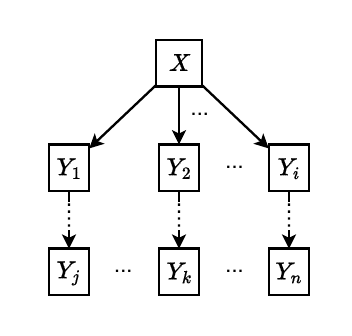}
		\caption{$Y_1,Y_2,\dots,Y_n$ are all the programs $X$ can be optimized into. $\{Y_1,Y_2,\dots,Y_n\}=\{Y|\exists \varphi\!\in\!\Phi^*\, Y\!=\!\varphi(X)\}$.}   
		\label{fig:flawXY}
	\end{subfigure}  
     \,\,
	\begin{subfigure}{0.49\textwidth}
		\captionsetup{skip=1pt}
		\centering
		\includegraphics[trim={0pt 14pt 0 0},clip, width=0.955\linewidth]{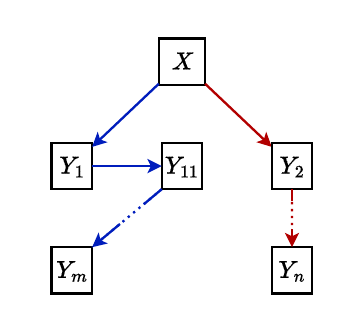}
		\caption{Two sequences of optimizations passes: $X \color{blue}\rightarrow \color{black}Y_1 \color{blue}\rightarrow \color{black} Y_{11} \color{blue}\rightarrow \cdots \color{black}Y_m$ and $X \color{red}\rightarrow \color{black}Y_2 \color{red}\rightarrow \cdots \color{black}Y_n$.}
		\label{fig:seqs}
	\end{subfigure} 
	\caption{The optimization of program $X$. Every box in the diagram represents a program while every arrow represent a optimization pass.
}
	\label{fig:flaw}
 \vspace{-5pt}
\end{figure}

\begin{theorem}[Solving the Phase Ordering Problem $\ne$ Finding the Globally Optimal Code]
\label{theo:subopt}
There exists a program $P$ for which the globally optimal code \wrt optimization phases $\Phi$ is not the same as the result of applying the optimal ordering of $\Phi$ to $P$.
\end{theorem}
\begin{proof}
To prove the above theorem, we first prove the following proposition, called \textit{\underline{L}ess \underline{E}fficient Programs Yield \underline{B}etter \underline{O}ptimizations (LEBO)}:

\vspace*{3pt}
\noindent
\textbf{Proposition} (LEBO). There exists a program $P$ for which there is another program $O$, such that (1) $O$ can be transformed into $P$ through a sequence of optimization passes (\ie, $\exists \varphi \in \Phi^{*}\, \varphi(O)=P$), and (2) applying the optimal ordering of $\Phi$ (for $O$) to $O$ results in a different program than applying the optimal ordering of $\Phi$ (for $P$) to $P$, \ie, $\varphi_O(O) \ne \varphi_P(P)$ where $\varphi_O$ and $\varphi_P$ denote the optimal ordering of $\Phi$ for $O$ and $P$ respectively. 
\vspace*{3pt}

Proof-by-contradiction: assume otherwise: there does \textit{not} exist a program $P$ for which there is another program $O$ such that $O$ can be transformed into $P$ via a sequence of optimization passes, and applying $\varphi_O$ to $O$ and $\varphi_P$ to $P$ respectively end up with different programs. This means that for any program $P$, every program $O$ that can be transformed into $P$ through a sequence of optimization passes will be optimized into the same program via $\varphi_O$ as $P$ is optimized into via $\varphi_P$. For presentation simplicity, we denote this postulation as $S$. 

Now, consider the setting illustrated in Figure~\ref{fig:flawXY}. Let $X$ denote a program, and $\mathcal{Y}$ denote the set of all programs that $X$ can be optimized into by some sequence of optimization passes, $\mathcal{Y} = \{Y|\exists \varphi \in \Phi^*\,Y = \varphi(X)\}$. This means for any program $Y_i \in \mathcal{Y}$, the relationship between $X$ and $Y_i$ aligns with that between $O$ and $P$ depicted in Figure~\ref{fig:flawOP}. According to postulation $S$, we can derive that applying the optimal phase ordering $\varphi_{Y_i}$ to any program $Y_i \in \mathcal{Y}$ results in the same program as applying the optimal phase ordering $\varphi_X$ to $X$. 

As illustrated in Figure~\ref{fig:seqs}, we apply randomly two distinct sequences of optimizations passes to $X$ until no more optimizations can be applied,\footnote{Considering the bounded code size, every sequence of optimization passes will be finite.} at which point we generate $Y_n$ and $Y_m$. Assume $Y_n$ differs from $Y_m$ and $\delta(Y_n) > \delta(Y_m)$ (this aligns with how programs are optimized in the real-world). 
According to the conclusion of the last paragraph, applying the optimal ordering $\varphi_{Y_m}$ to $Y_m$ results in the same program as applying the optimal phase ordering $\varphi_X$ to $X$. Since $\varphi_{Y_m}(Y_m) = Y_m$ and $Y_m$ is already less efficient than $Y_n$, applying $\varphi_{Y_m}$ to $Y_m$ does not result in the same program as applying $\varphi_X$ to $X$. Hence, the contradiction is derived, the Proposition LEBO holds. In fact, we can go a step further than Proposition LEBO. That is, since the programs that $P$ can be optimized into is a subset of the programs that $O$ can be optimized into, if $\varphi_O(O) \ne \varphi_P(P)$, it can only be the case that $\delta(\varphi_O(O)) > \delta(\varphi_P(P))$. Since there may exist another program $O^\prime$, such that $O^\prime \phiLequiv P$ and $\delta(O^\prime) > \delta(\varphi_O(O))$, the globally optimal code of $P$ \wrt $\Phi$ will be at least as efficient as $\varphi_O(O)$. In other words, the globally optimal code of $P$ \wrt $\Phi$ is guaranteed to be different from $\varphi_P(P)$. Therefore, we have proved the theorem.
\end{proof}


\subsection{The Relationship between the Problem of Finding the Globally Optimal Code \wrt Optimization Phases and the Phase Ordering Problem}
\label{subsec:rela}

While Theorem~\ref{theo:subopt} proves that the problem of finding the globally optimal code \wrt optimization phases is not equivalent to the phase ordering problem, it also indicates the specific condition under which the two problems become identical. That is, when there does not exist an alternative program to which applying the optimal phase ordering generates more efficient code than applying the optimal phase ordering to the input program, solving the two problems for such input programs is essentially the same. The reason this specific condition can hold is straightforward: if, for any program $P$, there exists a different program $O$, such that applying the optimal phase ordering to $O$ always results in a more efficient program than applying the optimal phase ordering to $P$, then there does not exist a globally optimal code for $P$, which contradicts with Theorem~\ref{theo:exi}. Theorem~\ref{theo:relation} formalizes the relation between the two problems.

\begin{theorem}[The relationship between the two problems]
\label{theo:relation}
The problem of finding the globally optimal code $Q$ for a program $P$ \wrt optimization phases \wrt $\Phi$ is finding a semantically equivalent program $O$ \wrt $\Phi$, such that applying the optimal ordering $\varphi_{O}$ of $\Phi$ to $O$ results in $Q$, \ie $\varphi_{O}(O) = Q$.
\end{theorem}
\begin{proof}
Given that $Q$ is the globally optimal code for $P$ \wrt $\Phi$, $P$ and $Q$ are semantically equivalent \wrt $\Phi$. According to Definition~\ref{def:equi}, there must exist a program $O$ such that $O$ can be optimized into $Q$ because either $P$ can directly be optimized into $Q$, or $P$ and $Q$ are semantically equivalent through other programs due to transitivity. In both cases $\exists O \exists\varphi\in\Phi^* \varphi(O) = Q$. Moreover, $Q$ must be the result of applying the optimal phase ordering $\varphi_{O}$ to $O$, otherwise, $Q$ will be less efficient than $\varphi_{O}(O)$, thus cannot be the global optimally code for $P$ \wrt $\Phi$.
\end{proof}

Based on the relationship between the two problems, we discuss the theoretical and practical advantages of solving the problem of finding the globally optimal code \wrt optimization phases.

\noindent\textbf{\textit{Generality from the Theoretical Perspective.}}\, Based on Theorem~\ref{theo:relation}, it should be clear that the problem of finding the globally optimal code \wrt optimization phases is a generalization of the phase ordering problem. Because if the program $O$ in Theorem~\ref{theo:relation} can be obtained by optimizing $P$ with $\Phi$, formally, $O \in \{P^\prime|\exists\varphi\in\Phi^* \varphi(P) = P^\prime\}$, then finding $Q$ for $P$ is the same for solving the phase ordering problem for $P$. Otherwise, if $O \notin \{P^\prime|\exists\varphi\in\Phi^* \varphi(P) = P^\prime\}$, finding $Q$ for $P$ requires finding $O$ first before applying $\varphi(O)$ to $O$.

\vspace{3pt}
\noindent\textbf{\textit{Insensitivity from the Practical Perspective.}}\, Theorem~\ref{theo:relation} also shows that the efficiency of the code resulted from applying the optimal phase ordering to the input program depends on the original form of that program. Among all programs that are semantically equivalent to the input \wrt optimization phases, applying the optimal phase ordering to some can yield more efficient code than applying it to others. This poses a challenge to software development practices. Because in real development settings, there are well-established code style guidelines that prioritize program readability and maintainability. These guidelines do not necessarily align with compilers' goal of generating the globally optimal code. Solving the problem of finding the globally optimal code would resolve this dilemma. Because no matter how programs are written, they will always be optimized into their global optimum. This means that the performance of software can be maximized without requiring developers to adhere to specific coding styles, thereby maintaining their productivity.For these reasons, we believe that finding the globally optimal code \wrt optimization phases is a critical problem we should strive to solve in the future.

\subsection{(Un)Decidability of the Problem of Finding the Globally Optimal Code}
\label{subsec:chara}


\begin{theorem}[Undecidability]
\label{theo:dec}
The problem of finding the globally optimal code for a program $P$ \wrt optimization phases \wrt $\Phi$ is undecidable.
\end{theorem}
\begin{proof}
To prove this theorem, we construct a reduction from the halting problem to the problem of finding the global optimum \wrt $\Phi$. For illustration purpose, we assume that $\Phi$ contains the necessary optimization passes for transforming the program in Figure~\ref{fig:ori1} to that in Figure~\ref{fig:comB1} (the reason is provide later in the proof).


    
    

\vspace*{3pt}
\noindent
\textbf{Step (1): construct a specific program}

Consider a program $P_h$ that we check for halting on an input $\alpha$. We construct a new program $P$ based on $P_h$. Specifically, we first include $P_h$ within $P$; then, after $P_h$ we add the following code that is independent from $P_h$. \myinline{var1} is a variable of integer type while \myinline{var2} is a boolean variable. For brevity, we ignore their declarations. 
\begin{lstlisting}[style=mystyle, basicstyle=\small\ttfamily\bfseries, numbers = none, xleftmargin=-4em,frame=none,framexleftmargin=1.5em, otherkeywords={printf}]
        printf("%d", ((var1/10) << 4) + var1 * var2 % 10);
\end{lstlisting}



\vspace*{3pt}
\noindent\textbf{Step (2): establish the dependency of the optimization of} \boldsymbol{$P$} \textbf{on whether} \boldsymbol{$P_h$}'s \textbf{halts on} \boldsymbol{$\alpha$}

Assuming there exists a property $\pi$ of $P_h$ such that if $P_h$ halts on $\alpha$, then $\pi$ is guaranteed to hold; otherwise, $\pi$ may be either \myinline{true} or \myinline{false}, depending on the input $\alpha$.
In the proof of Theorem~\ref{theo:nph}, we provide a concrete example of such a property $\pi$. We then assign \myinline{var2} the value of the property $\pi$ before the added code as follows:
\begin{lstlisting}[style=mystyle, basicstyle=\small\ttfamily\bfseries, numbers = none, xleftmargin=-4em,frame=none,framexleftmargin=1.5em, otherkeywords={printf}]
        var2 = @$\boldsymbol\pi$@;
        printf("%d", ((var1/10) << 4) + var1 * var2 % 10);
\end{lstlisting}

\vspace*{3pt}
\noindent\textbf{Step (3): reduce the halting problem to finding the global optimum of} \boldsymbol{$P$} \textbf{\wrt} \boldsymbol{$\Phi$}
Determining whether $P_h$ halts on $\alpha$ is in essence finding the globally optimal code $Q$ for $P$ \wrt $\Phi$. Because if $Q$ is found, we can derive from $Q$ if $P_h$ halted on $\alpha$. The reason is that $Q$ also contains the global optimum of the added code \wrt $\Phi$ since the added code is independent from $P_h$. The global optimum for the added code takes two forms. First, if $P_h$ halts on $\alpha$, the added code will be optimized into a code that is at least as efficient as 
\myinline{printf("
because, as explained earlier, $\Phi$ has the necessary optimization passes to perform the transformation. Otherwise, the added code 
will be optimized into a different global optimum that is less efficient than 
\myinline{printf("
because the value of \myinline{var2} cannot be determined definitively. Therefore, by examining which of the two global optimum of the added code is generated, we can decide if $P_h$ halts on $\alpha$. This means that solving the problem for finding the global optimum for $P$ would effectively solve the halting problem for $P_h$. Since the halting problem is undecidable, the problem for finding the global optimum must also be undecidable.
\end{proof}

Next, we prove the problem of finding the globally optimal code \wrt optimization phases $\Phi$ is also NP-hard.

\begin{theorem}[NP-hardness]
\label{theo:nph}
The problem of finding the globally optimal code for a program $P$ \wrt optimization phases $\Phi$ is NP-hard.
\end{theorem}
\begin{proof}
Similar to the proof of Theorem~\ref{theo:dec}, we aim to reduce the halting problem to the problem of finding the globally optimal code to show if we can find the global optimum efficiently (in polynomial time), we can solve halting problem, which is to known to be NP-hard, efficiently. 

Since we have already showed the reduction from the halting problem to the problem of finding global optimum, we only need to prove this reduction is polynomial-time computable. Recall the above-mentioned instance of the halting problem (\ie, whether a program $P_h$ halts on an input $\alpha$), the reduction consists of two parts: adding \myinline{printf("
to $P_h$ and finding the property $\pi$ of $P_h$. Since the former can obviously be done in constant time, we focus on the latter. 

One way to find a property $\pi$ of $P_h$, which only holds if $P_h$ halts on $\alpha$, is to instrument $P_h$ to create a property $\pi$. For instance, we can declare a \myinline{boolean} variable $\pi$ prior to the execution of $P_h$, which gets assigned the value \myinline{true} or \myinline{false}, depending on the input $\alpha$; we then find a program point $l$ in $P_h$ that will be executed \textit{iff} $P_h$ halts on $\alpha$. Essentially, $l$ can be anywhere within a basic block $B$ on the control flow graph (CFG) of $P_h$, such that $B$ is the only predecessor of the exit block, and the exit block is the only successor of $B$. If such basic block $B$ does not exist on the CFG of $P_h$, we can create it by linking all the predecessors of the exit block to $B$, and adding an edge to link $B$ to the exit block. At the program point $l$, we assign the value \myinline{true} to variable $\pi$. After this instrumentation of $P_h$, $\pi$ is guaranteed to be true \textit{iff} $P_h$ halts on $\alpha$. This procedure for instrumenting $P_h$ is a polynomial-time algorithm (potentially in number of predecessors of the exit block in the CFG of $P_h$). As mentioned earlier, adding the code \myinline{printf("
can be done in constant time. Therefore, our reduction from the halting problem to the problem of finding global optimum can be computed in polynomial-time.
\end{proof}

%% file: frame.tex
\section{A Conceptual Framework for finding the globally Optimal COde \wrt Optimization Phases}
\label{sec:frame}

This section presents a conceptual framework for finding the globally optimal code \wrt optimization phases, and a method for realizing this framework.

\subsection{A Conceptual Framework Based on Reverse Optimizations}
\label{subsec:frame}

We use \cmp to illustrate the key ideas of this conceptual framework which is capable of finding the global optimum for any program \wrt optimization phases.

Consider a program $P$ \cmp attempts to optimize by applying the optimal ordering of its optimization passes $\Phi$. The reason that \cmp may fail to generate the global optimum for $P$ \wrt $\Phi$ is, as explained previously, \cmp does not consider any program that is less efficient than $P$ during the optimization process. Therefore, it should be rather clear that augmenting $\Phi$ does not help \cmp generate the globally optimal code. In any case, Theorem~\ref{theo:subopt} still holds \wrt to the augmented set of optimization passes. A simple way to address this issue is to reverse each optimization pass $\phi \in \Phi$ and incorporate each reverse optimization to $\Phi$. With the new set of optimization passes, we show that \cmp is capable of finding the globally optimal code for any program.

\begin{theorem}[Capability of Finding the Global Optimum \wrt Optimization Phases]
\label{theo:capgo}
Let $\Phi$ denote a set of optimization passes, with each $\phi \in \Phi$ strictly increasing the efficiency of the generated code, \ie, $\delta(\phi(P)) > \delta(P)$ for an input program $P$. Let $\phi_{i}^{-1}$ denote the reverse of an optimization pass $\phi_{i} \in \Phi$, \ie, $\phi_{i}^{-1}(P_1) = P_2$ \textit{iff} $\phi_{i}(P_2) = P_1$. Let $\Phi^{-1} = \{\phi^{-1}|\phi \in \Phi\}$ denote the set of all reverse optimizations. With $\Phi \cup \Phi^{-1}$ (denoted by $\tilde{\Phi}$), a compiler is capable of generating the globally optimal code for any program.
\end{theorem}
\begin{proof}
For any program $P$, we first prove that the set of programs $\mathcal{P} = \{P^\prime|P^\prime \phiequiv P\}$ that is semantically equivalent to $P$ \wrt $\Phi$ does not change after the incorporation of $\Phi^{-1}$. This is because, for any program $P^\prime$ in $\mathcal{P}$, there does not exist a sequence of optimization passes $\tilde{\varphi} \in \tilde{\Phi}^{*}$ that can optimize 
$P^\prime$ into another program that is not already in the set $\mathcal{P}$. The reason is, no optimization pass $\tilde{\phi} \in \tilde{\Phi}$ can optimize any program in $\mathcal{P}$ to a program that does not belong to $\mathcal{P}$. If $\tilde{\phi} \in \Phi$, $\tilde{\phi}(P)$ is already in $\mathcal{P}$. Alternatively, if $\tilde{\phi} \in \Phi^{-1}$, then $\tilde{\phi}(P) \in \mathcal{P}$ because $\tilde{\phi}(P)$ can be optimized into $P$ via one optimization pass in $\Phi$, this means $\tilde{\phi}(P) \phiequiv P$. Since no optimization pass $\tilde{\phi} \in \tilde{\Phi}$ can discover any program outside of $\mathcal{P}$, no sequence of optimization passes $\tilde{\varphi} \in \tilde{\Phi}^{*}$ can do so either.

A compiler, with $\tilde{\Phi}$, is capable of generating the globally optimal code for any program $P$. The reason is the compiler can explore any program that is semantically equivalent to $P$ \wrt $\Phi$, since it can transform $P$ into any program $P^\prime \in \mathcal{P}$ with $\tilde{\Phi}$. Therefore, the theorem is proved.
\end{proof}

Under this framework, the key to finding the globally optimal code for a program $P$ is identifying which optimization passes should be reversed, because clearly, not all reverse optimizations are useful. For example, the reverse of deadcode elimination, which is inserting deadcode into programs, will not help generate the globally optimal code. Similarly, the reverse of function inlining, which extracts code into separate functions that can be called from various points in the program, also does not serve the purpose. Therefore, generating the globally optimal code \wrt optimization phases heavily depends on the problem defined as follows:

\begin{definition}[The Minimum Set of Optimization Passes to Reverse]
	\label{def:realpro}
    Let $Q$ denote the globally optimal code for a program $P$ \wrt a set of optimization passes $\Phi$. $\bar{\Phi}$ is the \textit{\underline{m}inimum set of \underline{o}ptimization passes to \underline{r}everse (MOR)} \textit{iff} $\bar{\Phi}$ is the smallest subset of $\Phi$ such that a sequence of optimization passes $\varphi \in (\bar{\Phi}^{-1} \cup \Phi)^{*}$ where $\bar{\Phi}^{-1} = \{\bar{\phi}^{-1}|\phi \in \bar{\Phi}\}$, which can transform $P$ into $Q$, exists. Formally, $\nexists \tilde{\Phi}\! \subsetneq\! \Phi\,\;\! \exists\varphi\!\in\!(\tilde{\Phi}^{-1} \cup \Phi)^{*} \, \varphi(P) = Q \wedge (|\tilde{\Phi}| < |\bar{\Phi}|)$.
\end{definition}

Once the MOR $\bar{\Phi}$ is discovered, to find the globally optimal code \wrt optimizations phases, we can focus on developing a strategy for identifying at each step
which optimization pass, either a normal optimization $\phi \in \Phi$ or a reverse optimization $\bar{\phi}^{-1} \in \bar{\Phi}^{-1}$, should be applied.

\subsection{A Method for Realizing the Conceptual Framework}
\label{subsec:frame}

To realize the proposed conceptual framework, we propose a simple yet intuitive method, called \textit{iterative bi-directional optimizations (IBO)}, aiming to generate more efficient code than existing optimization mechanisms (\eg, exhaustive search, GCC/LLVM at -O3).

\newlength{\textfloatsepsave} 
\setlength{\textfloatsepsave}{\textfloatsep}
\setlength{\textfloatsep}{14pt}
\begin{algorithm}[t]
	\algnewcommand\OP{\poc}
	\algnewcommand\prog{\ensuremath{P}\xspace}
	\algnewcommand\pset{\ensuremath{\mathcal{P}}\xspace}
	\algnewcommand\psetu{\ensuremath{\mathcal{O}}\xspace}
	\algnewcommand\poo{\ensuremath{P_t}\xspace}
	\algnewcommand\TS{\ensuremath{\mathcal{T}}\xspace}
	\algnewcommand\Ttmp{\ensuremath{\mathcal{T}_{tmp}}\xspace}
	\algnewcommand\TU{\ensuremath{\mathcal{T}_u}\xspace}
	\algnewcommand\step{\ensuremath{k}\xspace}
	\algnewcommand\len{\ensuremath{len}\xspace}
	\algnewcommand\DP{\pdc}
	\algnewcommand\tm{\ensuremath{\phi}}
	\algnewcommand\nso{\ensuremath{\tilde{\mathcal{O}}}\xspace} 
	\algnewcommand\nsoo{\ensuremath{\mathcal{O}^\prime}\xspace} 
 
	\algblockdefx[Foreach]{Foreach}{EndForeach}[1]{\textbf{foreach} #1 \textbf{do}}{\textbf{end foreach}}

	\caption{The detailed steps of \tool.}
	\label{alg:go}

    \hspace*{\algorithmicindent-26pt} \textbf{Input:} an input program $P$, a set of optimization passes $\Phi$, and the number of iterations $k$  \\
    \hspace*{-1\algorithmicindent-26pt} \textbf{Output:} the most efficient program $Q$ that \tool can generate 
	\begin{algorithmic}[1]
		\Procedure{IterativeBidirectionalOptimizations}{\prog, $\Phi$, \step}
        \State \psetu $\gets$ $\{\prog\}$, \nso $\gets \emptyset$, $Q \gets$ \prog
		\LineComment{  \;\;\; /* selecting some optimizations to reverse to keep the optimization process tractable */}        
        \State $\bar{\Phi} \gets$ \textit{SelectOptimizationsToReverse}($\Phi$)
		\For{$i$ \textbf{from} $0$ to \step} 
            \State \nso $\gets$ \Call{applyReverseOptimizations}{\psetu, $\bar{\Phi}$} 	\label{line:reverse}	
            \For{$O$ \textbf{in} \nso}
    	    \State $Q^\prime$ $\gets$ \textit{ApplyOptimalOrderingThroughExhaustiveSearch}($O$, $\Phi$) \label{line:optimalphase}
			\If{$\tm(Q^\prime) > \tm(Q)$}  \label{line:updategostart}
			\State  $Q \gets Q^\prime$ 
			\EndIf \label{line:updategoend}
		    \EndFor
            \State  $\mathcal{O} \gets \tilde{\mathcal{O}}$ \label{line:updateforreverse}
        \EndFor
        \State  \Return{\pd} 
		\EndProcedure
  
         \item[]
         
		\Function{applyReverseOptimizations}{\psetu, $\bar{\Phi}$}
        \State \nso $\gets$ $\emptyset$, \nsoo $\gets$ $\emptyset$
		\For{$O$ \textbf{in} \psetu}
		   \State $\nsoo \gets$ \Call{applyReverseOptimizationsToOneProgram}{$O$, $\bar{\Phi}$}
		   \State  \nso $\gets \nsoo \cup \nso$
	   \EndFor 
	   \State  \Return{\nso}
		\EndFunction

        \item[]

		\Function{applyReverseOptimizationsToOneProgram}{$O$, $\bar{\Phi}$}
        \State \nsoo $\gets$ $\emptyset$
		\For{$\bar{\phi}$ \textbf{in} $\bar{\Phi}$}
			\State $O' \gets \bar{\phi}^{-1}(O)$  \Comment{$\bar{\phi}^{-1}$ denotes the reverse of the optimization pass $\bar{\phi}$}
			\State  \nsoo $\gets \{O'\} \cup \nsoo$
	   \EndFor
	   \State  \Return{\nsoo}
		\EndFunction
	\end{algorithmic}
\end{algorithm}

At a high-level, \tool performs iterations of bi-directional optimizations, with each iteration consisting of (1) applying the reverse of some optimization passes to generate less efficient programs than the input; and (2) optimizing each generated program using the optimal ordering of all optimization passes discovered through exhaustive search. It is worth mentioning that \tool is not guaranteed to find the globally optimal code for any input program. Because, for one, \tool, which performs finite number of iterations, cannot explore the entire space of semantically equivalent programs to the input \wrt optimization phases; thus, may only discover a local optimum. In addition, for feasibility concerns, \tool only selects a few optimization passes to reverse from the technical standpoint (details are provided in Section~\ref{sec:eva}), and is likely to miss those in the MOR defined earlier. Despite all its shortcomings, \tool is a powerful optimization technique that frequently generates more efficient code than exhaustive search and the most aggressive optimization mode (\ie, -O3) in GCC/LLVM. 

Algorithm~\ref{alg:go} shows the detailed steps of \tool. It takes a program $P$, a set of optimization passes $\Phi$, and the number of iterations $k$ as inputs, and produces the most efficient program $Q$ it can find as output. Each iteration that \tool performs consists of the following steps: (1) \tool applies the reverse of a subset of $\Phi$ to $P$ (initially stored in $\mathcal{O}$ as the only program) to generate a set of programs $\tilde{\mathcal{O}}$ (line~\ref{line:reverse}). Then, (2) for each program $\tilde{O} \in \tilde{\mathcal{O}}$, \tool applies the optimal phase ordering to generate $O^\prime$ (line~\ref{line:optimalphase}). Next, (3) \tool updates $Q$, which saves the globally optimal code for $P$ \wrt $\Phi$ found so far. In particular, among all $O^\prime$ generated in step (2), if the most efficient program is more efficient than the global optimum $Q$ found so far, then $Q$ is updated to be that program; otherwise, $Q$ remains unchanged (line~\ref{line:updategostart}-\ref{line:updategoend}). Finally, (4) since $\mathcal{O}$ saves the set of programs to which reverse optimizations are applied in each iteration, \tool updates $\mathcal{O}$ to be $\tilde{\mathcal{O}}$ (line~\ref{line:updateforreverse}), and returns to step (1).

%% file: evav3.tex
\section{Evaluation}
\label{sec:eva}

This section evaluates the effectiveness of \tool.
In particular, we first describe in details our implementation of \FIBO{}.
Next, we present the results of our comprehensive evaluation of \tool, including a comparison with exhaustive search and state-of-the-art compilers --- GCC/LLVM. 

\subsection{The Implementation of Iterative Bi-Directional Optimizations}
We implement \FIBO{} within the LLVM infrastructure, which performs all transformations at the intermediate representation (IR) level --- transforming a given IR as input into a new IR as output. A key aspect of our implementation of \FIBO{} is selecting transformations to reverse. Since we have not solved the problem of identifying MOR, we 
design the following criteria to guide our selection of transformations from the technical standpoint.

\begin{itemize}[leftmargin=*]
    \item \textit{\textbf{Influentiality:}} As we explained throughout the paper, the key objective of applying the reverse of transformations is to generate less efficient programs, which hopefully can be optimized into a more efficient version than the original program can be optimized into. This means that, at the very least, the less efficient programs must be optimized via a different sequence of optimizations than the original program. In this regard, we choose transformations that have a significant influence on others, meaning, reversing them would modify a wide range of code constructs, consequently, enable the application of many other optimizations. One particularly effective strategy we devise is to reverse an optimization $\alpha$, the result of which obstructs another optimizations $\beta$. Then, the third optimization $\gamma$ over which $\beta$ takes higher precedence can be applied, initiating a new sequence of optimizations that is not possible on the original program (as $\gamma$ cannot be applied due to the higher priority given to $\beta$). An illustrative example is the \textit{Mem2Reg} optimization in LLVM. This optimization identifies loads and stores of local variables, converting them into SSA-form. If \textit{Mem2Reg} is reversed, optimizations such as \textit{InstCombine} (an arithmetic optimization in LLVM), which feeds off the results of \textit{Mem2Reg}, can be obstructed. When \textit{InstCombine} is not applied, the information of range/bits will be preserved, which means optimizations like \textit{SCCP} (Sparse Conditional Constant Propagation), can leverage more precise range information for propagating integer constants.

    \item \textit{\textbf{Diversity:}} Another way to increase the chances of applying different optimization sequences to the less efficient programs is to make less efficient programs themselves more diverse. \textit{SimplifyCFG} optimization in LLVM is an example that serves the purpose. The optimization is responsible for merging basic blocks, collapsing redundant branches, and others to simplify Control-Flow Graph (CFG). Naturally, the reverse of \textit{SimplifyCFG} produces CFGs with a great deal of variety; furthermore, \textit{SimplifyCFG} can be repeatedly reversed to keep diversifying the less efficient programs. As a result, other optimizations related to CFGs such as \textit{JumpThreading} (which eliminates jumps between basic blocks) can be enabled.

    \item \textit{\textbf{Simplicity:}} We prioritize optimizations that are straightforward to reverse. Reversing LICM (loop invariant code motion) is a good example. We simply put an expression evaluated outside of the loop into the loop where the result of the evaluated expression is used. In contrast, we avoid optimizations that are difficult to reverse. Take dynamic loop unrolling (DCU) as an example. DCU partially unrolls a loop with a certain number of repetitions determined at run-time. This optimization can split a loop into multiple loops before the unrolling, making reversal highly complex. To reverse DCU, we need to perform sophisticated semantic analysis to determine precisely the potential influence that one separate loop has on the others and vice versa. Furthermore, unifying the bodies of the separate loops --- especially if they have diverged after the split --- into a single operation over the induction variable is also challenging. Most critically, we must ensure that the total number of loop iterations remains consistent after reversing DCU. Due to these complexities, we do not attempt to reverse such intricate optimizations as DCU.
\end{itemize}

According to the three criteria above, Table~\ref{table:transforms} lists all the transformations we choose to reverse. These optimizations target a broad range of code constructs, such as CFG, memory operations, and arithmetic expressions, ensuring the effectiveness of our implementation of \FIBO{} in handing a wide variety of programs.
With the set of transformations selected for reversal, we implement \FIBO{} according to Algorithm~\ref{alg:go}, which takes a program, the number of iterations $k$ as input and produces an optimized program as output.

\setlength{\textfloatsep}{\textfloatsepsave}
\begin{table*}
    \captionsetup{skip=2pt}
    \caption{A list of reversed transformations.}
    \label{table:list}
    \centering
    \adjustbox{max width=1\textwidth}{
        \begin{tabular}{ | c | c | c|}
            \hline
            Optimization                                                                  & Description                       & Example of a reversal \\
            \hline
            \textit{InstCombine}                                                          & \tabincell{c}{Make instructions                           \\ more complex}&

            \begin{minipage}[m]{0.5\textwidth}
                \lstset{style=mystyle, numbers = none, xleftmargin=0.5em,frame=none, basicstyle=\small\ttfamily\bfseries}
\begin{lstlisting}[]
// Before       // After
int ret = a&&b; int ret = !(!a||!b);
\end{lstlisting}
            \end{minipage}
            \\
            \hline
            \textit{SimplifyCFG}                                                          & \tabincell{c}{Make control-flow                           \\more complex}  &
            \begin{minipage}[m]{0.5\textwidth}

                \lstset{style=mystyle, numbers = none, xleftmargin=0.5em,frame=none, basicstyle=\small\ttfamily\bfseries}
\begin{lstlisting}[]
// Before           // After
return c ? a : b;   if (c) return a;            
                    else return b;
\end{lstlisting}

            \end{minipage}
            \\
            \hline
            \textit{Reassociation}                                                        & \tabincell{c}{Reassociate expres-                         \\sions in a non-canoni- \\ calized order} &
            \begin{minipage}[m]{0.5\textwidth}

                \lstset{style=mystyle, numbers = none, xleftmargin=0.5em,frame=none, basicstyle=\small\ttfamily\bfseries}
\begin{lstlisting}[]
// Before             
int ret = a + b + c;  
// After
int ret = a + c + b;
\end{lstlisting}

            \end{minipage}
            \\
            \hline

            \tabincell{c}{\textit{LICM}                                                                                                               \\ (Loop Invariant \\ Code Motion)}               &  \tabincell{c}{Put loop-invariants\\ into loops}&    \begin{minipage}[m]{0.5\textwidth}

                \lstset{style=mystyle, numbers = none, xleftmargin=0.5em,frame=none, basicstyle=\small\ttfamily\bfseries}
\begin{lstlisting}[]
// Before
int a = 10, b = 20, c = a + b;
for(int i = 0; i < 100; i++) 
  cout << c;
// After
int a = 10, b = 2;
for(int i = 0; i < 100; i++) 
  cout << a + b;
\end{lstlisting}

            \end{minipage}
            \\
            \hline

            \tabincell{c}{\textit{Mem2Reg}                                                                                                            \\ (Promote memory\\ to register)}                      & \tabincell{c}{Demote register\\ to stack}&
            \begin{minipage}[m]{0.5\textwidth}

                \lstset{style=mystyle, numbers = none, xleftmargin=0.5em,frame=none, basicstyle=\small\ttfamily\bfseries}
\begin{lstlisting}[]
// Before   // After                       
return a;   int stack, *p = &stack;         
            *p = a;
            return *p;
\end{lstlisting}
            \end{minipage}
            \\


            \hline
            \textit{Loop Vectorizer}                                                      & \tabincell{c}{Scalarize the SIMD
            \\(Single Instruction/\\Multiple Data) \\ operations}  & \begin{minipage}[m]{0.5\textwidth}

                \lstset{style=mystyle, numbers = none, xleftmargin=0.5em,frame=none, basicstyle=\small\ttfamily\bfseries}
\begin{lstlisting}[]
// Before
__builtin_ia32_addps(result, a,  b);
// After
result[0] = a[0] + b[0];
result[1] = a[1] + b[1];
result[2] = a[2] + b[2];
result[3] = a[3] + b[3];
\end{lstlisting}

            \end{minipage}
            \\
            \hline

            \begin{tabular}{@{}c@{}}\textit{DSE}\,(Dead Store\\ Elimination)\end{tabular} & \tabincell{c}{Insert dead store                           \\ and move stores\\ around}                &
            \begin{minipage}[m]{0.5\textwidth}

                \lstset{style=mystyle, numbers = none, xleftmargin=0.5em,frame=none, basicstyle=\small\ttfamily\bfseries}
\begin{lstlisting}[]
// Before   // After                           
*p1 = a;    *p1 = arbitrary_value;
*p2 = 0;    *p2 = 0;
*p3 = 0;    *p1 = a;
            *p3 = 0;
\end{lstlisting}

            \end{minipage}
            \\
            \hline
        \end{tabular}}
    \label{table:transforms}
    \vspace{-8pt}
\end{table*}

\subsection{Experiments}
\subsubsection{Comparing \tool with Exhaustive Search}\label{subsubsec:dir}
While Theorem~\ref{theo:subopt} reveals the issue with solving the phase ordering problem in theory, it remains to be seen if \tool can actually generate more efficient code than the solution to the phase ordering problem --- exhaustive search --- in practice. We aim to find out the answer in this evaluation. 

\tool, which is realized within the LLVM infrastructure, also performs exhaustive search as part of its workflow. Therefore, for consistency and easy integration, we implement exhaustive search using LLVM (the latest release at the time of this evaluation, \ie, LLVM 19.1.2). Like \tool, exhaustive search works at the IR level. Since LLVM allows users to specify the ordering of optimization passes for a particular compilation through command lines, we provide all possible sequences of optimization passes enumerated in depth-first order as command line arguments. Each sequence of optimization passes saturates at the program to which no optimization passes can be applied,
\ie, the program is the result of applying that sequence of optimization passes to the input. 

\vspace{3pt}
\noindent\textbf{\textit{Test Programs.}}\,
We source test programs from two existing benchmarks: \textit{AnghaBench}~\cite{anghabench} and \textit{llvm-opt-benchmark}~\cite{opt-benchmark}. The former consists of the LLVM test suite and SPEC CPU\textsuperscript{\textregistered} 2006. The latter is an LLVM dataset for compiler optimization research, containing popular programs, such as Linux, QEMU, OpenSSL. Ensuring the significance of our evaluation, we focus on 12 well-established, highly influential programs, all of which have 10K+ stars on their Github repository (details are provided in the supplemental material). To reduce the complexity of our evaluation, we do not feed compilers a whole program. Instead, we use individual files of these 12 programs. Considering that exhaustive search is an exponential-time algorithm, we randomly pick 50 files of single function, each with no more than 100 lines of code, on which the exhaustive search is feasible to run. We believe these files should provide sufficient data points for this evaluation without significantly prolonging our experiments. We consider all optimization passes in the LLVM -O3 pipeline to implement exhaustive search. However, since we target single function files, we filter out inter-procedural optimizations like IPSCCP (Inter-Procedural Sparse Conditional Constant Propagation) and function inlining.

\vspace{3pt}
\noindent\textbf{\textit{Machines.}}\,
We compile and run all test programs using three machines (of x86 architectures) for all of our evaluations:
Intel\textsuperscript{®} Core™ i7-12700H, Intel\textsuperscript{®} Xeon\textsuperscript{®} Platinum 9242, and AMD Ryzen™ 9 7950X, each of which has 32GB memory and runs Ubuntu 22.04.2 LTS.

\vspace{3pt}
\noindent\textbf{\textit{Selecting the Optimal Code Generated by \tool and Exhaustive Search.}}\, To select the most efficient code generated by \tool or exhaustive search, we focus exclusively on performance, that is, among all compiled code generated by each sequence of optimization passes, we select the one that runs the fastest. To achieve this goal, we need to compile the entire program, rather than the individual files selected for this experiment. However, we must preserve the optimizations performed by \tool or exhaustive search on selected files. Thus, we first use \myinline{llc} to compile the IRs generated by \tool or exhaustive search for a selected file into many object files, each of which corresponds to one particular sequence of optimization passes. Then, we compile remaining files in the program using LLVM. Finally, we link each object file produced by \tool or exhaustive search with those produced by LLVM to create one executable file for the entire program. We run all executable files using test suites provided with the program, which achieve sufficient code coverage (\ie, over 90\% basic block coverage for every selected file). From all these executable files, we choose the executable file that on average runs the fastest across all test suites on every hardware architecture mentioned above. We note that, for every C/C++ file selected for this experiment, there exsits one executable file resulted from a particular sequence of optimization passes (identified by \tool or exhaustive search) that has the lowest average running time across all inputs on every target machine.

\begin{table}[t]
    \centering
    \captionsetup{skip=1pt}
    \caption{
    \#files shows the number of files for which \tool generates strictly more efficient code. Then, on these files specifically, the rest of columns compare the performance of exhaustive search and \tool. In those cells, the first value in a tuple deontes the result of exhaustive search while the second denotes the result of \tool.
    }
    \label{table:exhau}
    \begin{tabular}{|l|c|c|c|c|}
        \hline
        Comparisons & \#files & compile time (s) & \#instructions & running time (s)  \\
        \hline
        \begin{tabular}{@{}l@{}} Exhaustive search \\ vs. \tool ($k = 1$) \end{tabular}  &13 & (124.9, 679.1) & (21, 16) & (0.67, 0.53) \\
        \hline
        \begin{tabular}{@{}l@{}} Exhaustive search \\ vs. \tool ($k = 2$) \end{tabular}   &20 & (125.3, 4895.23) & (23, 15) & (0.71, 0.51) \\
        \hline
    \end{tabular}
\end{table}

\vspace{3pt}
\noindent\textbf{\textit{Results.}}\,
Table~\ref{table:exhau} presents the results of \tool compared to exhaustive search. In ``\#files'' columns, the value shows the number of files for which \tool generates strictly more efficient code, that is, \tool and exhaustive search generate the same code for the remaining files. We consider the code generated by \tool strictly more efficient if its average running time across all test suites on every hardware is shorter \textit{and} it contains fewer instructions in the IR compiled from the reduce file. The rest of the columns compare the performance of \tool and exhaustive search on these specific files (for which \tool generates strictly more efficient code), especially, in those cells, the first value in every tuple denotes the result of exhaustive search while the second value denotes the result of \tool. ``compile time (s)'' compares the average time spent by \tool and exhaustive search for compiling one file into LLVM IR. ``\#instructions'' compares the average number of instructions in the IR generated from the reduced file. ``running time (s)'' compares the average running time among every selected file, with each file measured by its average running time across all inputs on every architecture. 

The two rows show the performance of \tool with one and two iterations compared to exhaustive search. Given the high computational cost of \tool, it cannot afford additional iterations. Specifically, the number of programs \tool explores grows exponentially with the number of steps of reverse optimizations, in addition, the number of programs that a generated program can be optimized into also grows exponentially in the exhaustive search. For this reason, \tool (the second value in every tuple) is significantly slower than exhaustive search (the first value in every tuple) in terms of the compile time. However, even with one iteration, \tool finds significantly more efficient code. In particular, the IR generated by \tool contains on average 25\%+ few instructions, and runs on average 20\%+ faster. The margin increases with two iterations. This result shows that \tool can generate more efficient code than exhaustive search for real-world programs. Furthermore, it demonstrates the significant potential of our conceptual framework for finding the globally optimal code, considering that \tool is in fact a rather basic realization. Future, more effective instantiations, especially those can identify the MOR, should outperform exhaustive search by a wider margin.

\vspace{3pt}
\noindent\textbf{\textit{Examples.}}\, Figure~\ref{fig:ofslen} depicts a function for which \tool produced more efficient code than exhaustive search. The original function is extracted from QEMU, a widely used machine and userspace emulator and virtualizer. For brevity and clarity, we only show its reduced version (Figure~\ref{fig:oriex}). For this function, the optimization that should have been performed is to eliminate the body of the \myinline{if} clause from line 7 to 9 in Figure~\ref{fig:oriex}. Since the \myinline{if} condition at line 4 is equivalent to that at line 7, when the execution ever gets into the body of the \myinline{else} clause from line 7 to 11, the \myinline{if} condition at line 7 is guaranteed to be \myinline{false}. 

Exhaustive search did not find an ordering of LLVM's optimization passes that performs this optimization because it cannot find an ordering to derive the equivalence between the two \myinline{if} conditions. 
A natural question may arise: why LLVM does not have an optimization pass that can switch side for \myinline{len} in either \myinline{if} condition, which would have made the equivalence between the two \myinline{if} conditions obvious. The reason is, switching sides for \myinline{len} does not make the resultant code more efficient, therefore, would not be performed by any optimization pass. For example, the \myinline{if} condition \myinline{32 - len == ofs} at line 7 takes one instruction of subtraction and another of equivalence check. If we move \myinline{-len} from left to right, the generated code \myinline{32 == ofs + len} still takes one instruction for addition and another for equivalence check. Given that addition and subtraction instructions run equally fast on modern x86 architectures, the transformation proposed above will not be performed.

\begin{figure*}[t]
    \captionsetup{skip=5pt}
    \centering
    \begin{subfigure}{0.32\linewidth}
        \captionsetup{skip=1pt}
        \raggedright
        \lstset{style=mystyle, numbers = left, xleftmargin=2em,frame=single,framexleftmargin=1.5em,basicstyle=\footnotesize\ttfamily\bfseries}
\begin{lstlisting}[morekeywords={Object, String, bool},basewidth=0.52em]
void ori(void* ret, 
         unsigned ofs,
         unsigned len) {
  if (ofs + len == 32) {
    return;
  } else {
    if (32 - len == ofs) {
      // perform action 1
    } 
    else {
      // perform action 2
}}}
\end{lstlisting}
        \caption{}
        \label{fig:oriex}
    \end{subfigure}
    \;
    \begin{subfigure}{0.32\linewidth}
        \captionsetup{skip=1pt}
        \centering
        \lstset{style=mystyle, numbers = left, xleftmargin=2em,frame=single,framexleftmargin=1.5em,basicstyle=\footnotesize\ttfamily\bfseries}
\begin{lstlisting}[morekeywords={Object, String, bool},basewidth=0.52em]
void rev(void* ret, 
         unsigned ofs,
         unsigned len) {
  if ((ofs+len)-32==0) {
    return;
  } else {
    if ((-len + 32) +
        (-ofs) == 0) {
      // perform action 1
    } else {
      // perform action 2
}}}
\end{lstlisting}
        \caption{}
        \label{fig:rev}
    \end{subfigure}
    \;
    \begin{subfigure}{0.32\linewidth}
        \captionsetup{skip=1pt}
        \raggedleft
        \lstset{style=mystyle, numbers = left, xleftmargin=2em,frame=single,framexleftmargin=1.5em,basicstyle=\footnotesize\ttfamily\bfseries}
\begin{lstlisting}[morekeywords={Object, String, bool},basewidth=0.52em]
void opti(void* ret, 
         unsigned ofs,
         unsigned len) {
  if (ofs + len == 32) 
  {
    return;      
  }
  else 
  {
    // perform action 2
  }
}
\end{lstlisting}
        \caption{}
        \label{fig:opt}
    \end{subfigure}

    \caption{(\subref{fig:oriex}) depicts the original function; (\subref{fig:rev}), which is resulted from the reverse optimizations, can be optimized into a more efficient code (depicted in (\subref{fig:opt})) than (\subref{fig:oriex}) can ever be optimized into (which is itself).}
    \label{fig:ofslen}
    \vspace{-10pt}
\end{figure*}

Figure~\ref{fig:rev} shows the program \tool generated by applying the reverse of \textit{InstCombine} in Table~\ref{table:list} to the original function. Notably, exhaustive search within \tool finds an ordering of optimization passes for this function that removes the redundant \myinline{if} clause mentioned earlier. Specifically, (1) it first switches \myinline{-32} in the \myinline{if} condition \myinline{(ofs + len) - 32 == 0} (line 4) from left to right, because doing so would eliminate the instruction for subtracting \myinline{32}, and the resultant code takes one instruction for adding \myinline{ofs + len} and another for comparing \myinline{ofs + len} with \myinline{32}. Next, (2) it transforms \myinline{(-len + 32) + (-ofs) == 0} (line 7 to 8) into \myinline{32 == (len + ofs)} through the following sub-steps: reassociating \myinline{(-len) + (-ofs)} to \myinline{-(len + ofs)} so that the minus sign can be applied once to the result of \myinline{(len + ofs)} instead of twice to \myinline{len} and \myinline{ofs} separately; moving the entire expression \myinline{-(len + ofs)} to the other side, which completely eliminates the minus sign, resulting in \myinline{32 == (len + ofs)}. Finally, (3) it removes the dead \myinline{if} clause given that it recognizes the equivalence between \myinline{(ofs + len) == 32} and \myinline{32 == (ofs + len)}, thanks to the transformation performed in step (1) and (2). The optimized code is shown in Figure~\ref{fig:opt}. For interested readers, we have left a few more examples to the supplemental material.

\vspace{3pt}
\noindent\textbf{\textit{Examining the Improved Efficiency of Code Generated by \tool.}}\,
To understand deeper how the code generated by \tool is more efficient, we conduct a qualitative study.

First, given the different IRs generated by \tool and exhaustive search for a C/C++ file, we pinpoint the specific part of the \tool's IR fully responsible for the better optimization it performs. In particular, we develop a reduction technique, which systematically removes other parts of the original file that do not lead to discrepancies in the IR generated by \tool and exhaustive search. A criterion we design to guide the reduction procedure is that the contrast between the IRs generated by \tool and exhaustive search must remain consistent before and after the reduction. For example, if the IR generated by \tool contains 5 fewer (or more) \myinline{add} instructions than that generated by exhaustive search for an original file, then the IR generated by exhaustive search must also contain 5 fewer (or more) \myinline{add} instructions for the reduced file. We utilize the automatic test case reducer~\cite{regehr2012test} to repeatedly remove some code at a time, until no more code can be removed without violating the above-mentioned criterion. To ensure the validity of reduced programs, we use CompCert's interpreter~\cite{compcertUB} to detect and eliminate undefined behavior if they appear.

Next, we compile the final version of the reduced file into IRs with \tool and exhaustive search using the respective sequence of optimization passes that produces the most performant code for the original file. We then investigate how the IR generated by \tool for the reduced file is more efficient. For this manual inspection, we start with a comparison of the control flow graph (CFG) of the two IRs. If theirs CFG are identical, we then examine the differing sets of instructions. We find out almost always \tool's IR consists of fewer and more efficient instructions measured by LLVM cost model\footnote{\url{https://llvm.org/doxygen/CostModel_8cpp.html}}. If the CFGs differ, we analyze the discrepancies and discover that the improved efficiency is due to the simplified control flow structures, such as having fewer branches and basic blocks. Figure~\ref{fig:ofslen} is one example of the improved code efficiency stemming from the simplified CFG. Even within corresponding basic blocks that perform the same functionality in the two IRs, \tool's code is generally optimized more effectively, just as it is when the CFGs of the two IRs are identical.

\subsubsection{Improving the Optimizations of State-of-The-Art Compilers}\label{subsubsec:dir}

Since there does not exist an efficient algorithm for solving the phase ordering problem, state-of-the-art compilers, such as GCC and LLVM, employ their own optimization pipelines. Therefore, a natural and relevant question is can we utilize our conceptual framework to improve the effectiveness of GCC/LLVM's optimizations? To answer this question, we adapt \tool for this evaluation. In particular, we replace exhaustive search in Algorithm~\ref{alg:go} with GCC/LLVM's optimization scheme, while keeping the rest of the algorithm intact. We then compare the efficiency of the code generated by \tool and GCC/LLVM directly. If \tool generates more efficient code, we conclude \tool can be utilized to improve the optimization of GCC/LLVM. 

For both compilers, we use their latest release (14.2 for GCC and 19.1.2 for LLVM) at the time of this evaluation. We run -O3 optimization level for both GCC and LLVM, as it is the most aggressive optimization mode and likely to generate the most efficient code. Based on the adaption of \tool introduced above, we compare \tool against GCC and LLVM separately. When comparing \tool against a specific compiler, we configure \tool to run that compiler's -O3 optimization (\eg, \tool runs LLVM's -O3 when compared against LLVM). Running LLVM at -O3 within \tool is straightforward because \tool is implemented within the LLVM infrastructure. We simply run the \myinline{opt} tool at -O3 to optimize the IR that \tool produces after applying the reverse transformations. Integrating GCC into \tool is slightly more involved. We first convert the IR that \tool produces (with the reverse optimizations) into C files using \myinline{LLVM-CBE}. Then, we compile these files using GCC (at -O3) to generate the assembly code, which will be compared with the assembly code directly generated by GCC at -O3. 

We randomly pick 200 C/C++ files from the same set of programs used in the previous experiment, which should allow us to draw a sound conclusion from our evaluation results. In this experiment, we lift the restriction to single-function files, which is no longer necessary. We evaluate the efficiency of the code produced by \tool and GCC/LLVM as follows. First, we measure code size by counting the number of instructions in the generated code. In particular, when comparing \tool against LLVM, we count the number of instructions in the generated IR. When comparing against GCC, we count the average number of instructions in the assembly code generated by GCC and \tool (following the methodology described earlier) across the three target machines. To measure the running time of code generated by \tool, we adopt the procedure described in the previous experiment to make the code executable. Then, we record the average execution time of the code across all inputs on one hardware architecture. Last, this average execution time will be averaged again over all three architectures. For code generated by GCC/LLVM, we run the executable that they compiled from the entire program, and record the runtime for the selected files in the same manner described above. Finally, we perform the same qualitative evaluation to understand deeper about optimizations performed by \tool for each selected file. 

\begin{table}[t]
    \centering
    \captionsetup{skip=1pt}
    \caption{Compare \tool against GCC/LLVM.}
    \label{tab:gccllvm}
    \begin{tabular}{|l|c|c|c|c|}
        \hline
        Comparisons & \#files & compile time (s) & \#instructions & running time (s) \\
        \hline
        LLVM vs. \tool ($k = 1$)         & 49 &  (0.09, 0.59) & (32, 29) & (1.23, 1.04)       \\
        \hline
        LLVM vs. \tool ($k = 2$)         & 96 &  (0.08, 2.73) & (32, 28) & (1.18, 1.01)         \\
        \hline
        LLVM vs. \tool ($k = 3$)         & 109 & (0.08, 31.48) & (33, 26) & (1.34, 1.05)     \\
        \hline
        LLVM vs. \tool ($k = 4$)          & 114 & (0.07, 213.95)  & (31, 23) & (1.15, 0.92)    \\
        \hline
        LLVM vs. \tool ($k = 5$)          & 116 & (0.08, 1002.71) & (34, 20) & (1.37, 0.88)     \\
        \hline\hline
        GCC vs. \tool ($k = 1$)         & 40 &  (0.08, 0.62) & (33, 28) & (1.24, 1.06)      \\
        \hline
        GCC vs. \tool ($k = 2$)         & 99 & (0.06, 3.11) &  (31, 26) & (1.16, 1.03)       \\
        \hline
        GCC vs. \tool ($k = 3$)         & 114 & (0.07, 36.13) &  (32, 26) & (1.21, 1.06)        \\
        \hline
        GCC vs. \tool ($k = 4$)         & 119 &(0.07, 234.08) &  (31, 22) & (1.18, 0.96)        \\
        \hline
        GCC vs. \tool ($k = 5$)         & 121 & (0.08, 1420.15) &  (33, 21) & (1.29, 0.86)       \\        
        \hline
    \end{tabular}
    \vspace{-10pt}
\end{table}

\vspace{3pt}
\noindent\textbf{\textit{Results.}}\,
Table~\ref{tab:gccllvm} presents the results of \tool compared to GCC/LLVM. The columns denote the same meaning as in Table~\ref{table:exhau}. As the table shows, the number of programs for which \tool produces more efficient code than GCC/LLVM increases as the value of $k$ increases. By the time $k = 3$, \tool produces more efficient code than GCC/LLVM for more than half of the files. The margin continues to widen with further increases in $k$. In terms of the size of the code compiled from the reduced file and the running time of the code compiled from the original file, \tool's advantage over GCC/LLVM steadily grows as $k$ increases. When $k=5$, the number of instructions in the code generated by \tool is close to being only half of the instructions in the code generated by GCC/LLVM; in addition, the code generated by \tool runs 20\%+ faster than that generated by LLVM, and over 30\%+ faster than that generated by GCC. This shows the degree to which \tool can improve GCC/LLVM with reverse optimizations. On the other hand, we acknowledge the substantial computational cost due to these reverse optimizations, evidenced by the much longer compile time of \tool. However, we believe this issue can be effectively addressed in the future, especially with a more advanced realization of the conceptual framework that can efficiently identify MOR.

%% file: related.tex
\section{RELATED WORK}

This section surveys two streams of related work: phase ordering problem and superoptimizations.

\subsection{The Phase Ordering Problem}


Earlier work tackled the phase ordering problem by (1) proposing a compiler/linker design to allow optimization phases to be applied in any order and even repeatedly, thereby eliminating potential phase ordering issues~\cite{po1}; (2) using interactive compilation to enable developers to fine-tune optimization sequences for specific applications~\cite{po2, po3}.

Later, iterative compilation-based methods have gained popularity for solving the phase ordering problem~\cite{nobre2016graph, kulkarni2009practical, purini2013finding, kulkarni2006exhaustive}.~\citet{kulkarni2006exhaustive, kulkarni2009practical} showed that exhaustive evaluation could be performed in a reasonable time for most functions in a benchmark suite.~\citet{nobre2016graph} introduced an iterative method using a manually constructed graph based on compiler experts' understanding of how different compiler passes interact.
Recently, \citet{hellsten2023baco} introduced BaCO, a compiler autotuning method that leverages Bayesian optimization to iteratively generate optimized code.
~\citet{purini2013finding} search for the near-optimal optimization sequence by trying all the sequences in a pre-built set containing few good optimization sequences.

Recent work also incorporates machine learning techniques to explore the search space~\cite{ashouri2017micomp,cereda2020collaborative,chen2021efficient,AutoPhase,mammadli2020static,almakki2022autophase}. 
MiCOMP~\cite{ashouri2017micomp} employed neuro-evolution to create an artificial neural network that predicts effective optimization sequences.
Autophase~\cite{AutoPhase} use deep reinforcement learning to find a sequence of compilation passes that minimizes its execution time. 


\subsection{Superoptimization}

Unlike traditional compilers, which aim for partial optimizations, superoptimization involves exhaustively searching for the shortest and most efficient sequence of instructions within a specific instruction set architecture (ISA) to perform a given task~\cite{shypula2021learning, sasnauskas2017souper, albert2024superstack, xu2022quartz, schkufza2013stochastic}. 
STOKE~\cite{schkufza2013stochastic} treats the loop-free binary superoptimization task as a stochastic search problem to find an optimized version of a given target program according to two goals: maintaining correctness and improving performance. 
Souper~\cite{sasnauskas2017souper} is a synthesis-based superoptimizer for a domain-specific intermediate representation (IR) that resembles a functional, control-flow-free subset of LLVM IR.
SuperStack~\cite{albert2024superstack} superoptimizes stack-based bytecode via greedy, constraint-based, and SAT Techniques. 
Our method, unlike superoptimization, addresses the limited transformation issue of phase ordering, instead of finding the global optimum at ISA level.

%% file: conclusion.tex
\section{CONCLUSION}
In this paper, we challenge the long-standing focus on solving the phase ordering problem in compiler optimizations: solving the phase ordering problem may not result in the globally optimal code \wrt optimization phases. To address this issue, we propose a conceptual framework that is capable of finding the globally optimal code \wrt optimization phases in any program. The key idea behind this conceptual framework is to incorporate the reverse of existing optimization passes so that the framework can explore any program that compilers deem semantically equivalent to the input program. Within this framework, we show that the key to the problem of finding the globally optimal code is identifying the minimum set of optimizations to reverse, so that in the future, one can focus on formulating a policy of using such reverse optimizations to find the globally optimal code. As a preliminary effort, we realize this framework into a method, called \textit{IBO}. Our evaluation of \tool shows that it frequently generates more efficient code than exhaustive search, which is deemed a solution to the phase ordering problem. In addition, by incorporating reverse optimizations, \tool also outperforms state-of-the-art compilers --- GCC and LLVM. Given the significance of our findings, we expect our work to inspire new design principles for compiler optimization in the pursuit of generating the globally optimal code.  